\documentclass[fleqn,usenatbib]{mnras}

\usepackage{newtxtext,newtxmath}

\usepackage[T1]{fontenc}
\usepackage{ae,aecompl}

\usepackage{graphicx}	
\usepackage{amsmath}	
\usepackage{amssymb}	
\usepackage{gensymb}

\usepackage{xcolor}

\title[Methods for multiple telescope beam imaging and guiding in the near infrared]{Methods for multiple telescope beam imaging and guiding in the near infrared}

\author[N. Anugu et al.]{N. Anugu,$^{1,2}$\thanks{E-mail: narsireddy.anugu@fe.up.pt}
A. Amorim,$^3$
P. Gordo,$^3$
F. Eisenhauer,$^4$
O. Pfuhl,$^4$
M. Haug,$^4$
\newauthor{E. Wieprecht,$^4$}
E. Wiezorrek,$^4$
J. Lima,$^3$
G. Perrin,$^5$
W. Brandner,$^6$ 
C. Straubmeier,$^7$
\newauthor{J.-B. Le Bouquin,$^8$}
and P.J.V Garcia$^1$
\\
$^{1}$Faculdade de Engenharia, Universidade do Porto, rua Dr. Roberto Frias, 4200-465 Porto, Portugal;\\ CENTRA -- Centro de Astrof\'{\i}sica e Gravita\c c\~{a}o, IST, Universidade de Lisboa, P-1049-001 Lisboa, Portugal\\
$^{2}$School of Physics, Astrophysics Group, University of Exeter, Stocker Road, Exeter EX4 4QL, UK\\
$^{3}$Universidade de Lisboa - Faculdade de Ci\^{e}ncias,
Campo Grande, 1749-016 Lisboa, Portugal; \\ CENTRA -- Centro de Astrof\'{\i}sica e Gravita\c c\~{a}o, IST, Universidade de Lisboa, P-1049-001 Lisboa, Portugal\\ 
$^{4}$Max Planck Institute for extraterrestrial Physics, Giessenbachstr., 85748 Garching, Germany\\
$^{5}$Observatoire de Paris Meudon, 92195 Meudon Cedex, Paris, France\\
$^{6}$Max-Planck-Institut f\"ur Astronomie, K\"onigstuhl 17, 69117, Heidelberg, Germany\\
$^{7}$1. Physikalisches Institut, Universit\"at zu K\"oln, Z\"ulpicher Str. 77, 50937 K\"oln, Germany\\
$^{8}$Institut de Plan\'{e}tologie et d\'{}Astrophysique de Grenoble (IPAG), Grenoble, France
}

\date{Accepted XXX. Received YYY; in original form ZZZ}

\pubyear{2017}

\graphicspath{{figs/}}

\usepackage[none]{hyphenat}
\usepackage{amsmath}
\usepackage{import}
\usepackage{amsmath,latexsym,amsfonts}

\begin{document}
\label{firstpage}
\pagerange{\pageref{firstpage}--\pageref{lastpage}}
\maketitle

\begin{abstract}
Atmospheric turbulence and precise measurement of the astrometric baseline vector between any two telescopes are two major challenges in implementing phase referenced interferometric astrometry and imaging. They limit the performance of a fibre-fed interferometer by degrading the instrument sensitivity and astrometric measurements precision and by introducing image reconstruction errors due to inaccurate phases. A multiple beam acquisition and guiding camera was built to meet these challenges for a recently commissioned four beam combiner instrument, GRAVITY, at the ESO Very Large Telescope Interferometer. For each telescope beam it measures: a) field tip-tilts by imaging stars in the sky; b) telescope pupil shifts by imaging pupil reference laser beacons installed on each telescope using a $2 \times 2$ lenslet; c) higher order aberrations using a $9 \times 9$ Shack-Hartmann.  The telescope pupils are imaged for a visual monitoring while observing.  These measurements enable  active field and pupil guiding by actuating a train of tip-tilt mirrors placed in the pupil and field planes, respectively. The Shack-Hartmann measured quasi-static aberrations are used to focus the Auxiliary Telescopes and allow the possibility of correcting the non-common path errors between the Unit Telescopes adaptive optics systems and GRAVITY. The guiding stabilizes light injection into single-mode fibres, increasing sensitivity and reducing the astrometric and image reconstruction errors. The beam guiding enables to achieve astrometric error less than $50\,\mu$as. Here, we report on the data reduction methods and laboratory tests of the multiple beam acquisition and guiding camera and its performance on-sky. 
\end{abstract}



\section{Introduction}\label{sec:intro}

\begin{figure}
\centering
\def\svgwidth{\columnwidth}
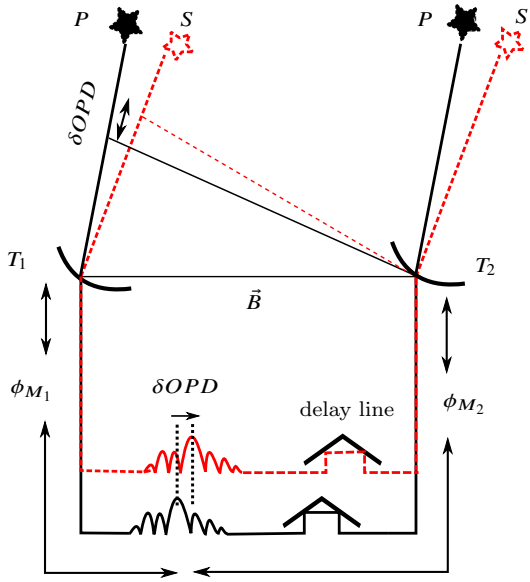
\caption{GRAVITY working principle. The beams of science $S$ and reference $P$ stars are combined simultaneously. The GRAVITY metrology system measures the differential optical path between the two stars introduced by the VLTI beam relay and the GRAVITY beam combiner instrument. The $\phi_{M_1}$ and $\phi_{M_2}$ are the phases between the beam combiner and the telescopes ($T_1$ and $T_2$). $\vec{B}$ is the baseline vector.}
\label{fig:fig1}
\end{figure}

GRAVITY~\citep{GRAVITYCollaboration2017} is a dual feed phase referencing interferometric imaging and a narrow-angle astrometric instrument. It has been built for the ESO Very Large Telescope Interferometer (VLTI)  and was born with the aim to monitor stellar sources in the vicinity of the Galactic Centre supermassive black hole~\citep{Genzel2010,Eisenhauer2011}. It combines four  beams coherently in the K-band of either Unit Telescopes (UTs, 8\,m) or Auxiliary Telescopes (ATs, 1.8\,m). It is designed to deliver a differential astrometry of around  10\,microarcseconds ($\mu$as) and an angular resolution of around 4\,milliarcseconds (mas). It works by observing a reference star at position $\vec{P}$ and a science star at position $\vec{S}$: fringes for each star are obtained in separate beam combiners (cf. Figure~\ref{fig:fig1}).    The differential separation of stars in the sky is proportional to  the differential optical path difference  ($\delta OPD$) observed in the beam combiner instrument. For first-order approximation, the differential optical path difference is related to the differential separation of stars and the baseline vector ($\vec{B}$) by

\begin{equation*}
\delta OPD  = ({\vec{S}-\vec{P}}) \cdot \vec{B}.
\end{equation*}

The coherent beam combination of star light is implemented using a single-mode fibre-fed integrated optics chip~\citep{Benisty2009, Jocou2014}. The $\delta OPD$ between the science and reference beam combiners is measured with a dedicated laser metrology system~\citep{Lippa2016, GRAVITYCollaboration2017}. 

\begin{figure}
\centering
\def\svgwidth{\columnwidth}
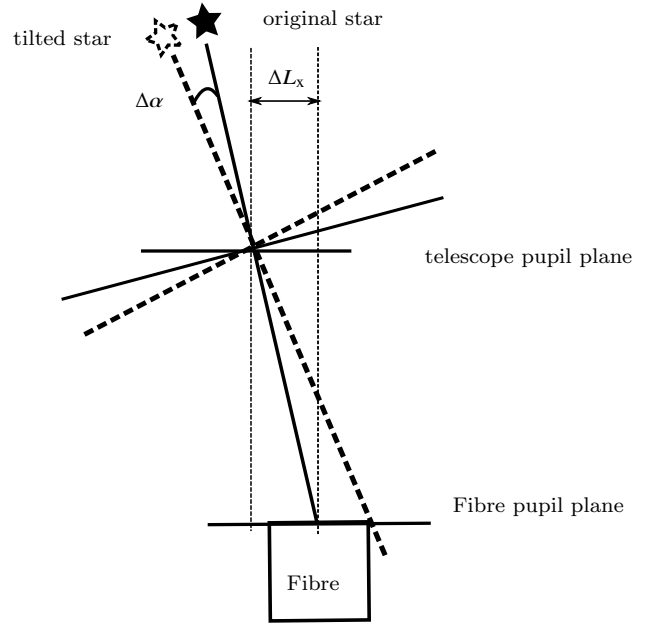
\caption{Influence of field drift and pupil lateral shift  in the beam while injecting light into a single-mode fibre. The star shown in dashed line is tilted ($\Delta \alpha$) with respect to the original star (solid line). The telescope pupil is shifted laterally ($\Delta L_{\rm x}$) with respect to the fibre pupil.}
\label{fig:fig2}
\end{figure}

Adaptive optics correction residuals such as slow field drifts (tip/tilts) and non-common path aberrations and telescope/optical pupil motions\footnote{Optical train vibrations cause pupil motions while tracking the object of interest with the delay lines.} (lateral and longitudinal or pupil defocus shifts) are a concern to be corrected. They degrade the performance in terms of: a) instrument sensitivity, because of star's light injection problems into the single-mode fibres in the presence of field drifts;   b) astrometric errors due to  field drifts and pupil shifts caused baseline paradox~\citep{Lacour2014,  Woillez2013, Colavita2009};  c) image reconstruction errors due to inaccurate phases\footnote{Of relevance for, e.g., imaging stellar high signal-to-noise surfaces.}.

The astrometric error caused by the presence of  field tip-tilts ($\Delta \alpha$), lateral  ($\Delta L_{\rm x}$) and longitudinal ($\Delta L_{\rm z}$) pupil shifts   is ~\citep[cf. Figure~\ref{fig:fig2}, ][]{Lacour2014}

\begin{equation}\label{sigma_T}
\sigma_{\rm T} \sim \dfrac{\Delta L_{\rm x} \sin(\Delta \alpha)   +  \Delta L_{\rm z} \big[1-\cos(\Delta \alpha)\big]   }{B} \hspace{0.5cm} \mathrm {rad}.
\end{equation}   

In order to achieve the GRAVITY goal of 10\,$\mu$as astrometric precision, the telescope beams must be corrected with residuals in the field to less than 10\,mas root-mean-square (RMS) deviation error\footnote{This is given by $\sim 0.2\lambda_\mathrm{K}/D$, with $\lambda_\mathrm{K}=2.2\,\micron$, $D=8$\,m (Lacour et al. 2014).}, the lateral pupil to less than 40\,mm RMS and the longitudinal pupil to less than 10\,km RMS, at the UTs~\citep{Lacour2014}. To enable these corrections a four beam imaging and guiding instrument, the multiple beam acquisition and guiding camera, was built.

\begin{figure*} 
\centering
\def\svgwidth{0.95\textwidth}
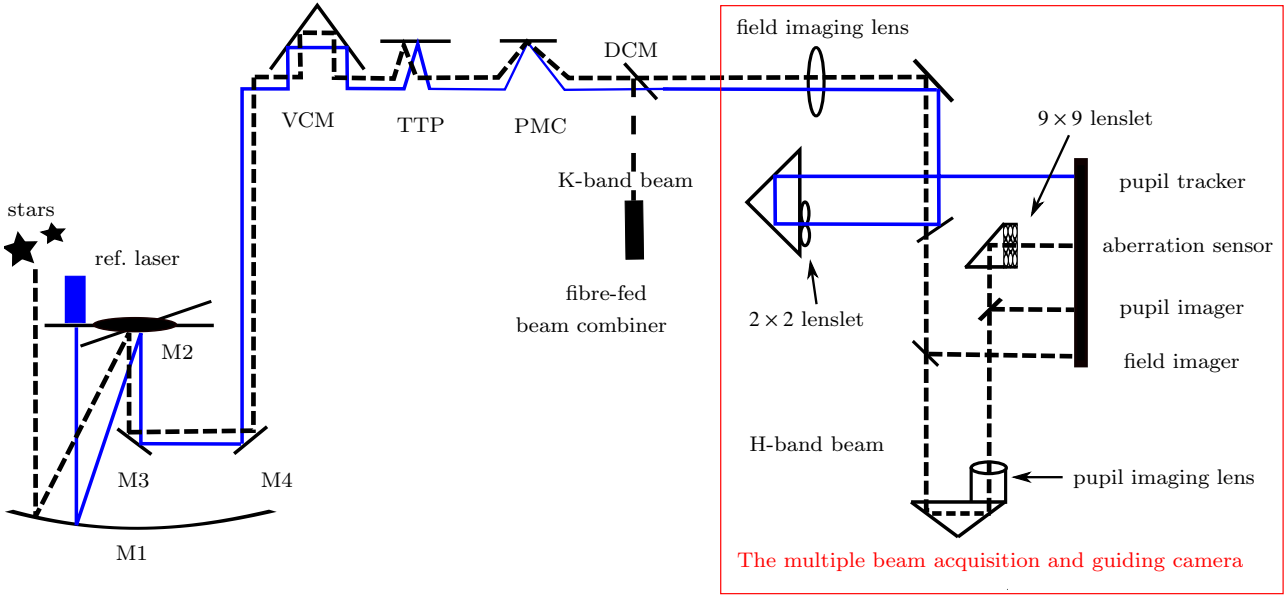
\caption{The multiple beam acquisition and guiding camera conceptual scheme. For simplicity, only one telescope beam case is presented. Only the H-band (dashed line) and $1.2\,\micron$ wavelength (emitted from reference laser beacon, in solid line and blue colour) are used for beam analysis. The K-band (wide spaced dashed line) is used for interferometric observations. A dichroic mirror (DCM) splits the incoming telescope beam and sends: a) the reflected light into the fibre-fed beam combiner; b) the transmitted light H-band and $1.2\,\micron$ wavelength beams into the multiple beam acquisition and guiding camera (red box in the Figure). The camera optical functions: a) pupil tracker; b) aberration sensor;  c)  pupil imager; and d) field imager.  The optical functions are used to stabilize the incoming beam and to inject the K-band beam into the fibre-fed beam combiner using variable curvature mirror (VCM), tip-tilt and piston mirror (TTP) and pupil motion controller (PMC) actuators.}
\label{fig:fig3}
\end{figure*}

The paper is presented as follows. In Section~\ref{sec:acq}, the optical layout and functions of the multiple beam acquisition and guiding camera are briefly described. The methods of extraction of the beam guiding parameters from the image are presented in Section~\ref{sec:method}. Section~\ref{sec:lab_char} and~\ref{sec:sky_char} present the instrument concepts  validation and  characterization results at the laboratory and on-sky.  The paper ends with a summary and conclusions.

\section{The multiple beam acquisition and guiding camera optical functions}\label{sec:acq}
\subsection{Overall optical layout}

The multiple beam acquisition and guiding camera is a subsystem of GRAVITY. It has two sub-units: a) the external folding optics; and b) the core beam analyser. The folding optics has four optical entrance channels, accepts four 18\,mm sized near~infrared beams and redirects them into the  beam analyser. The beam analyser is the core of the multiple beam acquisition and guiding camera. It implements four important optical functions (cf. Figure~\ref{fig:fig3}): a)  pupil tracker -- measures the telescope pupil shifts; b) aberration sensor -- measures quasi-static wavefront aberrations; c) pupil imager -- images the telescope pupil; and d) field imager -- images the field and measures slow field drifts.

The multiple beam acquisition and guiding camera images are used to stabilize the telescope beams using tip-tilt and piston systems (TTPs),  pupil motion controllers \citep[PMCs,][]{Pfuhl2014} and variable curvature mirrors \citep[VCMs,][]{Ferrari2003}. The tip-tilt and piston systems are actuated mirror controllers placed in the pupil plane of the optical train where incoming beams are propagated. These are used to correct the field drifts. The pupil motion controllers are the pupil lateral shift mirrors located in a field plane of the optical train. They are used to correct the lateral pupil offsets.  The variable curvature mirrors are  focus/defocus mirrors placed at the image plane in the VLTI delay lines~\citep{Derie2000}. These are used to adjust the position of pupil longitudinally (pupil focus correction). 


The specifications of the multiple beam acquisition and guiding camera have been defined as a trade-off between the beam guiding requirements (see Table~\ref{table:GRAVITYSpecs}) for the key Galactic Centre astrometry program of GRAVITY and the technical constraints~\citep{Amorim2012}. The camera is designed to work in H-band wavelengths (1.45-1.85\,$\micron$) for the following reasons: a)~the Galactic Centre is situated behind $\sim$\,30 magnitudes of visible extinction~\citep{Genzel2010} and is brighter in the near~infrared; b)~the VLTI throughput is optimized in the infrared as  most of the visible light is fed into the visible wavefront sensors at the transmitted Coud\'e focus~\citep{Arsenault2003}; c)~the H-band is as close as possible to the science wavelengths (K-band). To increase stability and to reduce thermally emitted backgrounds, the camera operates at cryogenic temperatures (the detector at 80\,K,   the beam analyser at $\sim 110$\,K, the external folding optics at 240\,K). The core optical functions of the camera are manufactured with a single optical material (fused silica, the coefficient of thermal expansion $\sim 10^{-6}\,^\circ\mathrm{C}^{-1}$) to minimize stresses when it is cooled down to the cryogenic  temperature. The optical functions  are imaged on a $2048 \times 2048$~pixel Hawaii-2RG detector~\citep{Finger2008, Mehrgan2016}.  The detector is operated in correlated double sampling readout mode with the frame rate of 0.7\,s and the readout noise of $13\,\mathrm{e}^-\,\mathrm{pix}^{-1}$ RMS error. The optical functions of the beam analyser are detailed in the following sections. 

\begin{table}
	\centering
	\caption{UT beam guiding requirements for a $m_\mathrm{H}=13$\,mag star~\citep{Lacour2014}.}
	\label{table:GRAVITYSpecs}
	\begin{tabular}{l*{3}{c}r}
		\hline
		Parameter & Value (RMS)&   \\
		\hline
		Lateral pupil guiding  & $\leq 40$\,mm  \\
		Longitudinal pupil guiding &  $\leq 10$\,km  \\
        Field guiding & $\leq 10$\,mas \\ 
        Wavefront accuracy & $\leq 80$\,nm \\
		\hline
	\end{tabular} 
\end{table}

\subsection{Pupil tracker}\label{sec:PTdesign}

The VLTI pupils are re-imaged via delay lines in the GRAVITY beam combiner instrument.  In between the telescope pupil and the GRAVITY fibre-fed beam combiner there exist several optical pupils including the delay lines. The pupils move laterally and longitudinally, due to the optical train vibrations, while tracking the object of interest. The GRAVITY beam combiner obtains the effective pupil, i.e. the barycentric mean of all these randomly positioned pupil apertures in the whole optical train from the telescope to the beam combiner. The effective pupil is different from the telescope pupil. The typical maximum lateral pupil error is around $\sim 5$\% of the pupil ($\sim 0.4$\,m for an 8\,m pupil), much larger than the 40\,mm required to achieve  $10\,\mu$as astrometry  (see Table~\ref{table:GRAVITYSpecs}).

While equalizing the optical paths of the telescopes with the delay lines, the longitudinal pupil position (or pupil defocus) has to be adjusted constantly using the delay line variable curvature mirrors to preserve the field of the VLTI~\citep[$2^{\prime \prime} \times 2^{\prime \prime}$,][]{Ferrari2003}. The variable curvature mirror corrects the longitudinal pupil positions (pupil defocus) by adjusting its curvature based on a calibration pointing model. Limitations in the pointing model introduce residual longitudinal pupil drifts. The typical maximum longitudinal pupil error is around 30\,km at the UT. Whereas the requirement for the $10\,\mu$as astrometry is 10\,km (see Table~\ref{table:GRAVITYSpecs}).  In order to measure these pupil shifts, the pupil tracker was designed ~\citep{Pfuhl2012, Amorim2012}.  

To detect the pupil motions, the telescope pupil is imaged by a $2 \times 2$ lenslet. Telescope pupil reference laser beacons (1.2\,$\micron$) are used. For each telescope, four beacons are mounted on the four symmetric spider arms (cf. left of Figure~\ref{fig:fig4}).   The barycentre of these four laser beacon positions gives the centre of the pupil. These telescope pupil reference laser beacons are separated by 1.6\,m and 0.4\,m diagonally on the UTs and ATs, respectively (cf. Table~\ref{table:PTSpecs}).

\begin{figure}
\centering
\def\svgwidth{\columnwidth}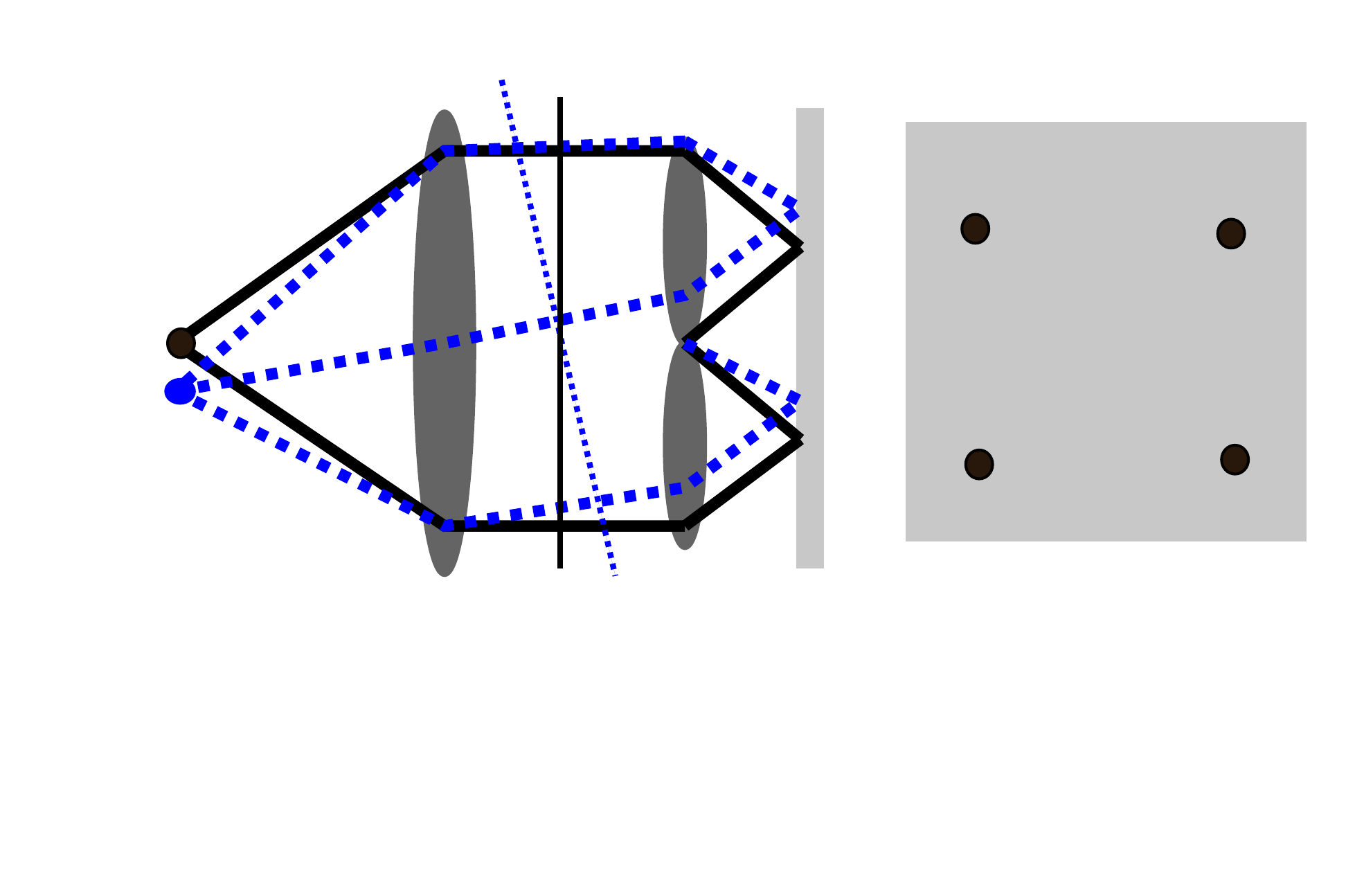
\def\svgwidth{\columnwidth}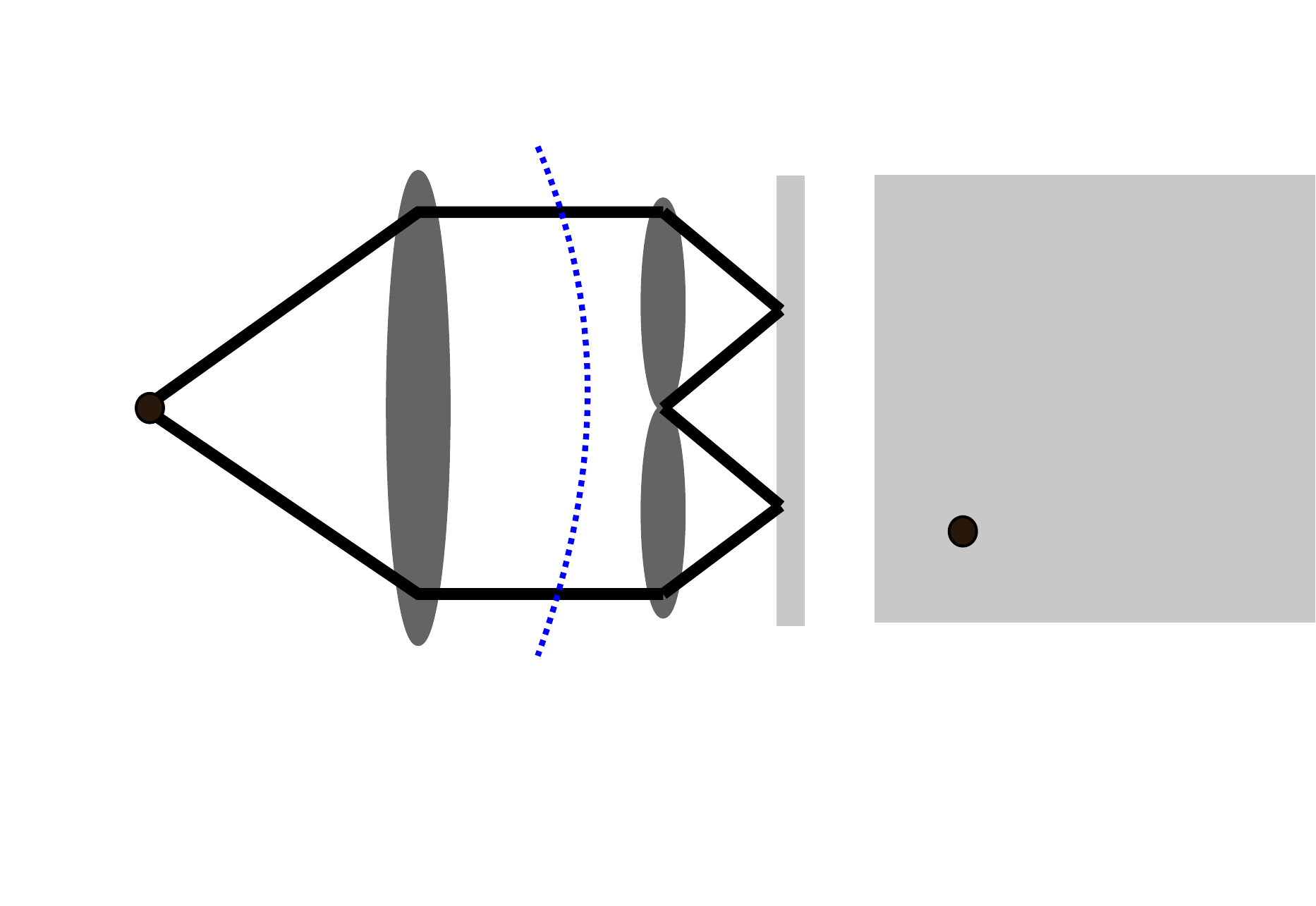
\caption{Pupil tracker working principle.  The reference telescope pupil and corresponding laser beacon spots are represented by the solid lines in black and  filled circles respectively.  The shifted pupil and corresponding telescope pupil reference laser beacon spots are represented by dashed lines in blue and star symbols respectively. A typical lateral shifted pupil is presented in panels of (a) and (b). A typical longitudinal shifted pupil is presented in panels of (c) and (d).}
\label{fig:fig5}
\end{figure}

Figure~\ref{fig:fig5}\,(a) and (b) illustrate a case of a laterally shifted pupil, i.e., the 2 $\times$ 2 lenslet formed spots are laterally shifted from the reference telescope pupil formed spots.  Figure~\ref{fig:fig5}\,(c) and (d) illustrate a case of longitudinally shifted pupil, i.e., the lenslet formed spots have either converged ($L_{\rm z}<0$) or diverged ($L_{\rm z}>0$) with respect to the reference spots depending on the direction of the longitudinal displacement. 

\begin{table}
	\centering
	\caption{Pupil tracker specifications for the UTs.}
	\label{table:PTSpecs}
	\begin{tabular}{l*{3}{c}r}
		\hline
		Parameter & Value &   \\
		\hline
		Wavelength & 1.2\,\micron  \\
		$2 \times 2$ lenslet &  on $2.03 \times 2.03$\,mm  \\
        Focal length ($f_\mathrm{PT}$) & 14\,mm \\
        Reference beacons separation& 1.6\,m\\
        Field of view  & 11.2\,m   \\
		Pixel scale  & 59.86\,mm\,pixel$^{-1}$ \\
        Detector pixel size ($d_{\rm p})$ & 18\,\micron \\
        Field stop & $2^{\prime\prime}\times 2^{\prime\prime}$\\
		\hline
	\end{tabular} 
\end{table}

\subsection{Pupil imager}

The pupil imager is designed to verify the pupil image quality visually (e.g. to identify pupil vignetting or pupil shifts). It is used to align GRAVITY with the VLTI during the commissioning/installation and visually monitor the telescope pupil during observations.    It images the telescope pupils using telecentric lenses. Its specifications are presented in Table~\ref{table:PISpecs}.

\begin{table}
	\centering
	\caption{Pupil imager specifications for the UTs.}
	\label{table:PISpecs}
	\begin{tabular}{l*{3}{c}r}
		\hline
		Parameter & Value &   \\
		\hline
		Wavelength & H-band (1.45-1.85 \micron) \\
		Field of view  & 11.2\,m\\
		Pixel scale  & 54\,mm\,pixel$^{-1}$ \\
        Flux for $m_\mathrm{H}=13$\,mag & $1.8\times 10^4~\mathrm{e}^{-}\,\mathrm{s}^{-1}\,\mathrm{image}^{-1}$ \\
        Field stop & $2^{\prime\prime}\times 2^{\prime\prime}$\\
		\hline
	\end{tabular} 
\end{table}

\subsection{Field imager}

The UTs are equipped with near~infrared wavefront sensors based adaptive optics ~\citep[CIAO,][]{Scheithauer2016}. The ATs, currently are only equipped with tip-tilt sensors and they will have adaptive optics modules in the near future~\citep[NAOMI,][]{Gonte2016}.  In both cases, GRAVITY experiences slow field drifts greater than 10\,mas due to its correction residuals and the VLTI tunnel seeing effects. The field imager aims to track these slow field motions. Its specifications are presented Table~\ref{table:FISpecs}.  The slow field tracking (with a frequency of 1\,Hz) is to be complemented by a fast tip-tilt monitoring system. This system works by injecting the visible tip-tilt laser beacon beams before the VLTI tunnel and by imaging them inside GRAVITY, using the position sensitive diodes~\citep{Pfuhl2014}. It will measure the field motions between the telescopes and GRAVITY.

\begin{table}
	\centering
	\caption{Field imager specifications for the UTs.}
	\label{table:FISpecs}
	\begin{tabular}{l*{3}{c}r}
		\hline
		Parameter & Value &   \\
		\hline
		Wavelength & H-band (1.45-1.85 \micron) \\
		Field of view  & $4^{\prime \prime} \times 4^{\prime \prime}$ \\
		Pixel scale  & 17.78\,mas\,pixel$^{-1}$ \\
        Flux for $m_\mathrm{H}=13$\,mag star & $4.8\times 10^4~\mathrm{e}^{-}\,\mathrm{s}^{-1}$ \\
		\hline
	\end{tabular} 
\end{table}

\begin{figure} 
\centering
\def\svgwidth{\columnwidth}
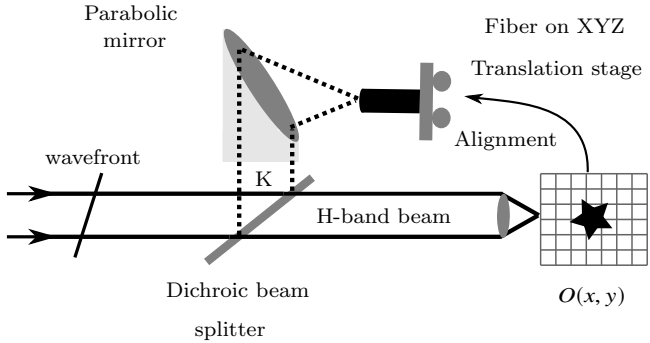 
\caption{Star light injection into a single-mode fibre.}
\label{fig:fig6} 
\end{figure}

In GRAVITY,  single-mode fibres feed light into the integrated optics ~\citep{Jocou2014,GRAVITYCollaboration2017}, which implements the actual beams combination.  The light injection into these fibres is stabilized by using the field imager measured field drifts (cf. Figure~\ref{fig:fig6}). The accurate position of star $O(x,y)$ is detected by the field imager using the H-band beam. Using this position, the single-mode fibre is aligned to the parabola focus with the fibre positioner~\citep{Pfuhl2014}.  Also using this position, the star light injection into the single-mode fibre is stabilized using the tip-tilt mirror located in front of the incoming beam.

\subsection{Aberration sensor}

The aberration sensor is designed to detect the quasi-static residuals of the UTs adaptive optics  system~\citep{Scheithauer2016}. For the ATs, it measures the telescope static aberrations as the ATs are not equipped with the adaptive optics yet. The measured quasi-static aberrations are used, once in a while, to focus the ATs or to correct the non-common path aberrations between the UTs adaptive optics system and GRAVITY.  It consists of four Shack-Hartmann sensors, each with $9 \times 9$  sub-apertures. Its specifications are presented in Table~\ref{table:ABSSpecs}.     

\begin{table}
	\centering
	\caption{Aberration sensor specifications for the UTs.}
	\label{table:ABSSpecs}
	\begin{tabular}{l*{3}{c}r}
		\hline
		Parameter & Value    \\
		\hline
		Wavelength & H-band (1.45-1.85 \micron) \\
		Lenslet &  $9 \times 9$    \\
		Field of view per sub-aperture & 4$^{\prime\prime}$\\
		Pixel scale  & 250\,mas\,pixel$^{-1}$ \\
		Sub-aperture size & $16 \times 16$~pixel$^2$ \\
        Maximum Zernike order measured & 28 \\
        Sub-aperture flux for an $m_\mathrm{H}=13$\,mag star & $760~\mathrm{e}^{-}\,\mathrm{s}^{-1}$  \\
        Field stop & $2^{\prime\prime}\times 2^{\prime\prime}$\\
		\hline
	\end{tabular} 
\end{table}

The pixel scale magnification in the field between the UTs and the ATs is $8/1.8 = 4.44$, i.e. the field imager 17.78\,mas/pixel scale at the UTs becomes 79\,mas for the ATs. The pixel scale de-magnification in between the UTs and ATs for the lateral pupil and longitudinal pupil  is $1.8/8 = 1/4.44$ and $(1.8/8)^2 = 1/4.44^2$, respectively.

\section{Methods for beam guiding parameters extraction}\label{sec:method}

\subsection{Image acquisition and analysis software}

A dedicated real-time software extracts the beams stabilization parameters from the detector images. The software is implemented within the ESO control software framework. It is written in C and C++ using mainly the ESO standard Common Library for Image Processing \citep[CLIP,][]{Ballester2008} and Common Pipeline Library \citep[CPL,][]{McKay2004} and installed in the VLT common software. It works on a Linux operating system based instrument workstation and does the image analysis in three steps. Firstly, every 0.7\,s it integrates the multiple beam acquisition and guiding camera detector image and stores it in the instrument workstation shared memory. Secondly, it copies the detector image from the instrument shared memory and by analysing the input detector image it evaluates beams tip-tilts, pupil shifts and beam aberrations. These  parameters are written to the instrument database and used for the beam stabilization in the closed loop. The multiple beam acquisition and guiding are carried out in real-time on the instrument in parallel to the observations, for all four telescopes.

\subsection{Pupil tracking}\label{sec:PTmes}

\begin{figure}
	\centering
	\def\svgwidth{0.45\columnwidth} 
	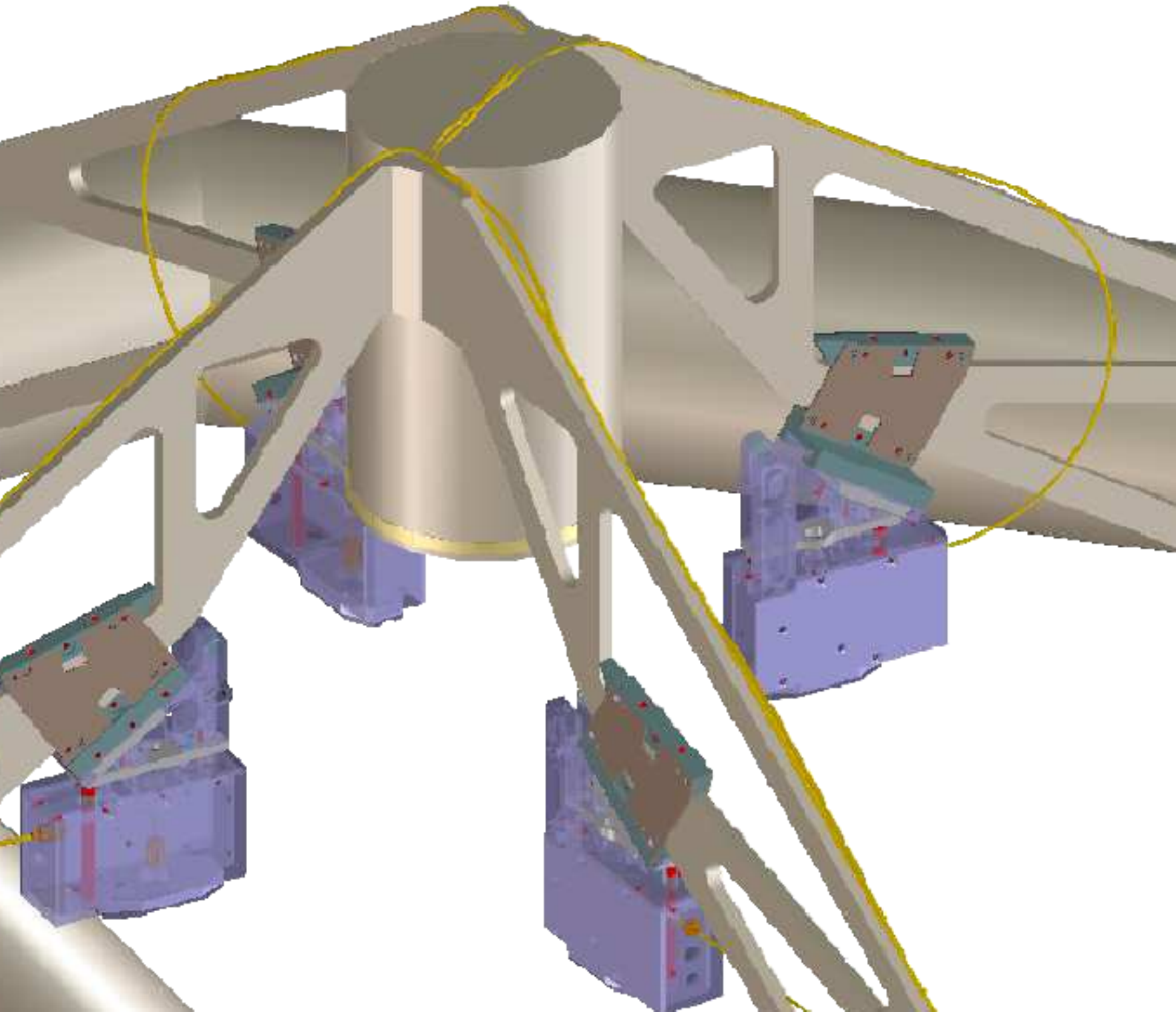 
	\def\svgwidth{0.5\columnwidth} \parbox[t]{1pt}{\rotatebox{90}{\hspace{1.5cm} meter}} \hspace{5pt}
   \includegraphics[width=0.45\columnwidth]{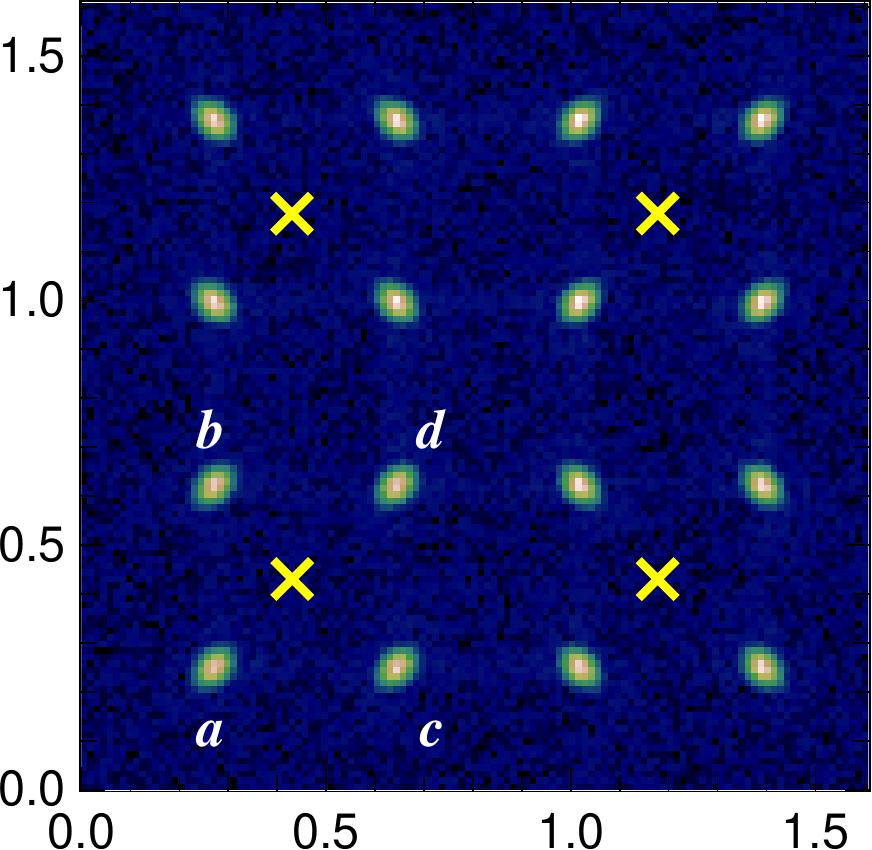} 
   
 \hspace{5cm}meter
 \caption{Left: external pupil reference beacons (blue boxes) installed on the telescope secondary mirror M2 spiders.  These telescope pupil reference beacons  are fed by a common laser diode via multi-mode fibres. Right: A numerically simulated pupil tracker image. The $a$-$b$-$c$-$d$ spots are the pupil reference beacons for a sub-aperture. The cross-marks $\times$ in the middle  are the centres of the lenslet sub-apertures.} 
    \label{fig:fig4}
\end{figure}

The lateral and longitudinal shifts of the telescope pupils are translated as tip-tilt and defocus aberrations when imaged on the 2 $\times$ 2 lenslet. The pupil shifts are evaluated by analysing the pupil tracker image in three steps. First, the reference $2 \times 2$ sub-aperture spots are generated by illuminating an internal reference laser source, which is implemented by the calibration unit~\citep{Blind2014}. Next, each sub-aperture $j$ position ($R^\mathrm{x,y}_j$) is computed by applying Gaussian fits. Second, the telescope pupil reference laser beacons spots in the detector image (see Figure~\ref{fig:fig4}) are identified using a flux threshold  ($\sim 5\sigma$ above background) and their full-width-half-maximum (FWHM) thresholds. Next, their positions are computed. From the spot positions,  the centre of each sub-aperture  ($C^\mathrm{x,y}_{j}$) is determined by taking the average of the telescope reference laser beacon positions.  Finally, the spot shifts ($S^\mathrm{x,y}_{ j}$) are computed by comparing them with the internal reference sub-aperture positions ($R^\mathrm{x,y}_{ j}$)
\begin{equation*}
S^\mathrm{x,y}_{ j} =  C^\mathrm{x,y}_{ j} - R^\mathrm{x,y}_{j}.
\end{equation*}

The telescope lateral pupil shifts are evaluated by taking the average of these spot shifts along the $X$ and $Y$ directions 

\begin{equation}\label{eq:Lxx}
 L_\mathrm{x,y}= \frac{1}{4} \sum_{j=1}^4 S_{j}^\mathrm{x,y}, 
\end{equation}

The longitudinal pupil shifts are evaluated by computing the amount of convergence or divergence (in radians) of the spot shifts~\citep[Eq.~\ref{eq:Lzz},][]{Salas2005}

\begin{equation}\label{eq:Lzz}
 L_\mathrm{z}= \frac{d_\mathrm{p}}{f_\mathrm{PT}}
 \frac{(S_{2}^\mathrm{x}+ S_{3}^\mathrm{x})
 -(S_{1}^\mathrm{x} + S_{4}^\mathrm{x}) 
 + (S_{1}^\mathrm{y}+  S_{2}^\mathrm{y})
 - (S_{4}^\mathrm{y} + S_{3}^\mathrm{y})}{8},  
\end{equation}

\noindent where $f_{\rm PT}$ is the focal length  of the $2 \times 2$ lenslet and $d_{\rm p}$ the detector pixel size (cf. Table~\ref{table:PTSpecs}).

\subsection{Pupil imaging}\label{PImes}

The pupil imager is used to visually monitor the telescope pupils for their quality during the preparation for observations and also while observing.  However, it can also be used, in future, for the lateral pupil tracking in the presence of a bright astrophysical source. A cross-correlation algorithm~\citep{Poyneer2003} allows computing the pupil shift from a shifted pupil image  by correlating it with the reference pupil image.\footnote{The reference pupil image is the pupil image where one wants to lock the pupil during observations. It can be a long exposure image of the telescope pupil.}

\subsection{Object tracking with the field imager}\label{FImes}

The field imager measures these field drifts by tracking the brightest object in the user-specified region (window) of interest. Windowing is important in crowded regions such as the Galactic Centre. The position of the brightest object is computed in two steps. Firstly, stars in the field are scanned using a $5\sigma$ flux threshold above background and a predetermined FWHM. Next, the detected stars are sorted in decreasing flux order. Secondly, the accurate position of the brightest object $O(x,y)$ is computed with a Gaussian fit in a window size of $8\times8$ pixel$^2$ centred on the object.  To track binaries where both stars are of comparable magnitude, the position of the second brightest object within the window is also measured, the tracking reference is chosen by the instrument operator.

The atmospheric differential refraction effects  are of relevance  because the H-band photo-centre is used to estimate the fibre coordinates to inject the K-band beam into the fibre. The shift ($\Delta R$) caused by the atmospheric differential refraction~\citep[e.g.][]{Roe2002} is  

\begin{equation}\label{ADR_shift}
 \Delta R  \simeq 206265^{\prime\prime}\,  \bigg(\dfrac{n_{\lambda_{\rm K}}^2-1}{2n_{\lambda_{\rm K}}^2}-\dfrac{n_{\lambda_{\rm FI}}^2-1}{2n_{\lambda_{\rm FI}}^2}\bigg)\tan \zeta,
\end{equation}

\noindent with $\lambda_\mathrm{FI}$ and $\lambda_\mathrm{K}$ the field imager (H-band) and the fibre (K-band) accepting wavelengths,  $n_{\lambda_\mathrm{FI}}$ and $n_{\lambda_\mathrm{K}}$ the corresponding refractive indices and  $\zeta$ is the object zenith distance. The refractive index for a given wavelength and in function of the observing conditions (temperature, pressure and humidity) is computed using Eq.~1 in \citet{Roe2002}.   

Since the effective wavelength of the field imager, $\lambda_{\rm FI}$, changes with the colour of the object, its value is modelled~\citep{Stone2002} as a function of the H-K colour considering the atmospheric transmission and the instrument transmission\footnote{Product of all filters transmission, reflection of mirrors and efficiency of the detector as a function wavelength.}. 

The atmospheric differential refraction object offset ($\Delta R$) correction is applied in real time given the H-K object colour inserted by the observer in the observing block.

\subsection{Beam wavefront sensing with the aberration sensor}\label{ABmes}

The reconstruction of the aberrated wavefront is done by comparing the reference Shack-Hartmann spot locations with the aberrated spot locations. The reference spots were obtained as part of the calibration procedure using a collimated beam. During the observations, the  spots of incoming telescope beams are acquired. The centroids of these spots are computed by applying  Gaussian fits. When an extended scene is observed, the centroids are computed using a cross-correlation algorithm~\citep{Poyneer2003}. Slopes of the spot shifts are computed by taking differences between the target and the reference spot positions. These slopes are normalized by  the focal length of the lenslet. Using these slopes, higher order aberrations of the input beams are evaluated in terms of Noll's Zernike coefficients~\citep{Dai1996} up to 28 in number.

\subsection{End-to-end simulations}

A meticulous numerical modelling of the multiple beam acquisition and guiding camera  is carried out to verify the beam analysing methods. These  simulations are implemented in the \texttt{Yorick}\footnote{\href{http://yorick.sourceforge.net/}{http://yorick.sourceforge.net/}} programming language for an easy interpretation, debugging and plotting. The LightPipes~\citep{Vdovin1997} ANSI C library was ported to \texttt{Yorick} and the adaptive optics \texttt{yao} \citep{Rigaut2013} library was used.  

The multiple beam acquisition and guiding camera detector images are generated using the optomechanical model parameters~\citep[in Zemax,][]{Amorim2012} and by applying the VLTI laboratory experimentally measured field motions, pupil motions (lateral and longitudinal) and higher order wavefront aberrations.  Finally, the input beam aberrations are reconstructed back and characterised  the multiple beam acquisition and guiding camera performance in terms of its accuracy in the presence of adverse imaging conditions and sensitivity towards the faint stars. Table~\ref{table:GRAVITY_end_to_end_specs} presents the numerical analysis of the accuracy of the  multiple beam acquisition and guiding camera.  

\begin{table}
	\centering
	\caption{End-to-end simulation of the guiding parameters accuracy and comparison with the requirements.}
	\label{table:GRAVITY_end_to_end_specs}
	\begin{tabular}{lcc}
		\hline
		Parameter & Requirement  &  Simulation\\
        & (RMS)  &  (RMS)\\
		\hline
		Lateral pupil guiding  & $\leq40$\,mm  & $\leq4$\,mm \\
		Longitudinal pupil guiding & $\leq10$\,km  & $\leq200$\,m\\
        Field guiding & $\leq10$\,mas & $\leq2$\,mas \\
        Wavefront measurement & $\leq80$\,nm  &$\leq70$\,nm \\
		\hline
	\end{tabular} 
\end{table}

\section{Laboratory characterization}\label{sec:lab_char}

\begin{figure}
\parbox[t]{1pt}{\rotatebox{90}{\hspace{-2.1cm}meter}} 
\parbox[t]{1pt}{\rotatebox{90}{\hspace{-6.5cm}arcsec}} 

\centering	
\includegraphics[width=0.45\textwidth]{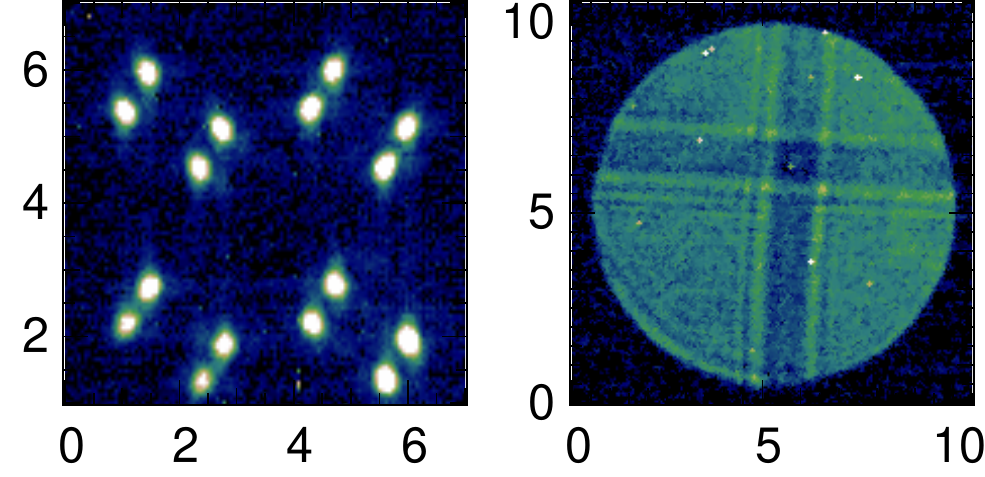} meter \hspace{3cm} meter
\includegraphics[width=0.45\textwidth]{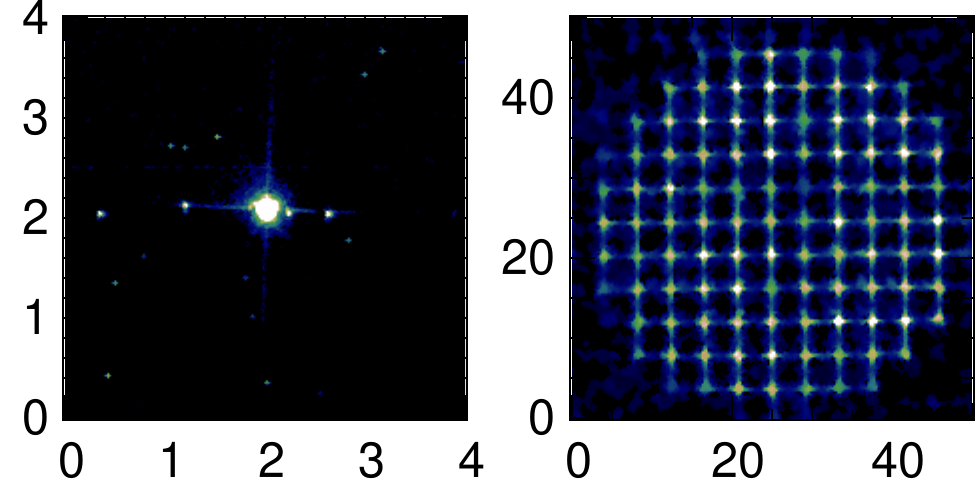} arcsec \hspace{3cm} arcsec
\caption{Optical imaging modes of the  camera obtained using the calibration unit generated beams. Images are scaled to the UT magnification. The top panel is the  pupil tracker and pupil imager from left to right, respectively. The bottom panel is the  field imager and aberration sensor imager from left to right, respectively.  The complex structure in the pupil image is laboratory specific~\citep{Blind2014}.} 
\label{fig:fig7}
\end{figure}

The multiple beam acquisition and guiding camera is characterized using the GRAVITY calibration unit in the MPE laboratory~\citep{Blind2014}.  The calibration unit simulates an artificial launch telescope designed by a parabolic mirror and two artificial stars with 95\% Strehl ratio by illuminating two single-mode fibres. The four telescope pupil reference laser beacon beams are generated by four multi-mode fibres inserted in a pupil plane.  Figure~\ref{fig:fig7} presents the imaging modes of the  camera, for one telescope, when illuminated by the calibration unit.

\begin{figure}
\centering     
\parbox[t]{11pt}{\rotatebox{90}{\hspace{0.5cm} Normalized flux}} 
\includegraphics[width=0.3\columnwidth]{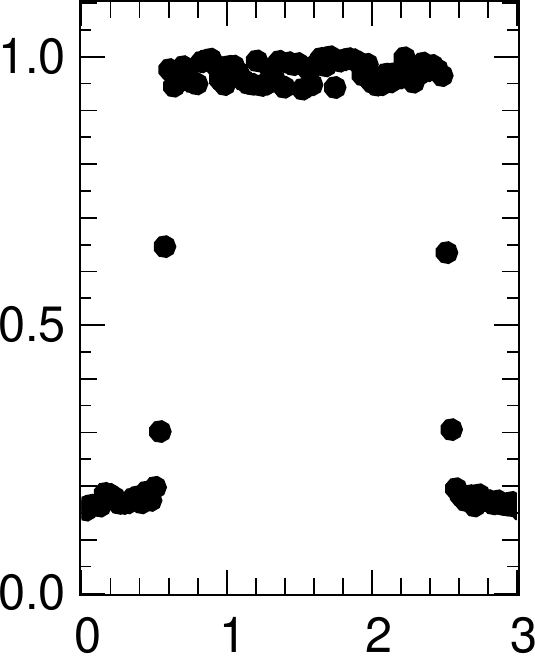}   \includegraphics[width=0.3\columnwidth]{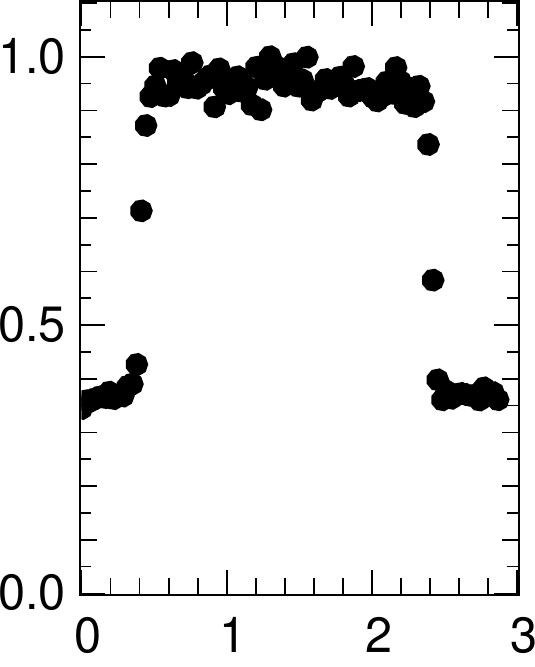}  
\includegraphics[width=0.3\columnwidth]{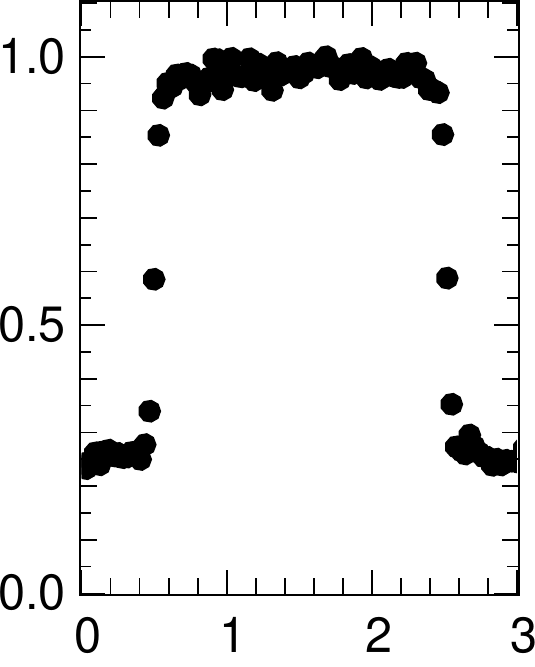}  

\hspace{1cm} FoV($^{\prime\prime}$) \hspace{1.5cm} FoV ($^{\prime\prime}$) \hspace{1.5cm} FoV ($^{\prime\prime}$)
\caption{Target normalized flux in a function of position for the  pupil tracker, pupil imager and aberration sensor windows (from left to right).}
\label{fig:FoVPupil}
\end{figure}

The field of view (FoV) of the pupil tracker, pupil imager and aberration sensor are estimated by scanning (applying tilts) a star target over their field of view. During the scanning, the flux and position of the  target are recorded in each window. Figure~\ref{fig:FoVPupil} presents the flux in the pupil tracker, pupil imager and aberration sensor windows as a function of star target position.  The field stop of each optical function is apparent in the width of the curves in Figure~\ref{fig:FoVPupil}, being $\simeq 2^{\prime\prime}$. The field stops as built  matched  the design values (cf. Tables~\ref{table:PTSpecs},\ref{table:PISpecs} and \ref{table:ABSSpecs}). 


\begin{figure}
\parbox[t]{1pt}{\rotatebox{90}{\hspace{0.8cm} $\Big|L_{\rm {x0}} - L_{\rm x}\Big|$ (mm) }}\hspace{5pt}
\centering
\hspace{3pt}
\includegraphics[width=0.45\columnwidth]{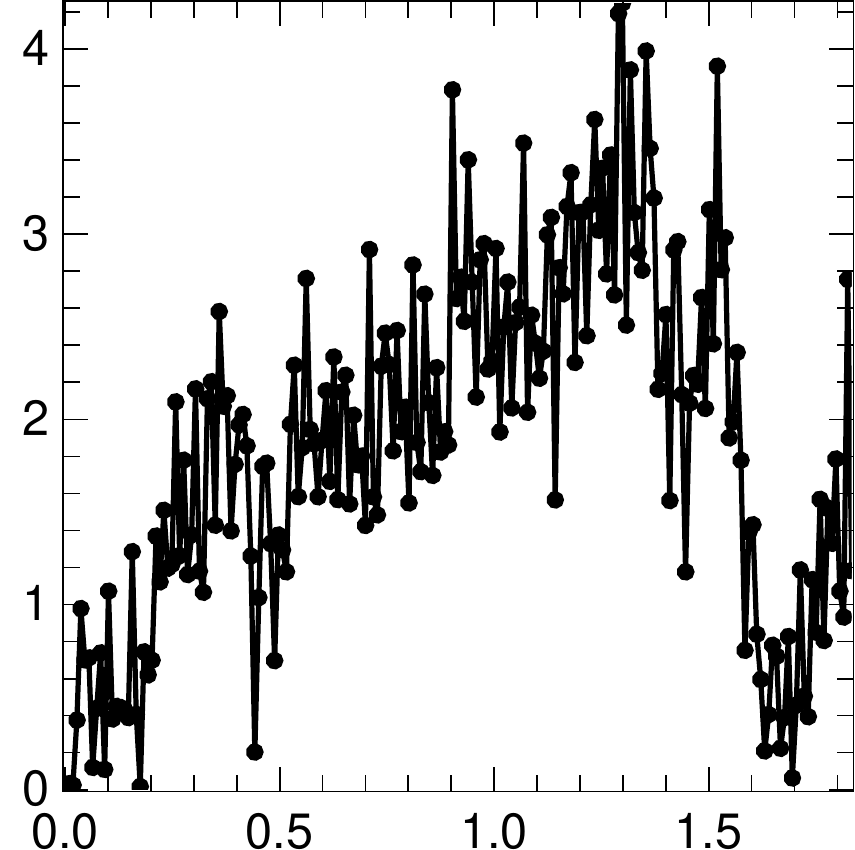} 

\hspace{1cm}$L_{\rm {x0}}$\caption{Absolute error in the lateral pupil position  measurement (UT beam) as a function of input lateral pupil shift in the laboratory. The error bars are RMS errors.} 
\label{fig:fig8}
\end{figure}

\subsection{Pupil tracking  performance}

The pupil tracker pixel scale is measured by equating the separation  of telescope pupil reference laser beacons observed on the  detector to the separation of those  beacons installed on the telescope of the calibration unit. The measured value is within 5\% of the design value of (Table~\ref{table:PTSpecs}).

The lateral pupil tracking performance is characterized in the laboratory by manipulating the pupil lateral correcting actuators. This experiment is carried out in two steps. First, known lateral pupil shifts are applied to the incoming beams by actuating the pupil lateral correcting actuators. Second, the input pupil shifts are estimated from the pupil tracker images as explained in Section~\ref{sec:PTmes}.  Figure~\ref{fig:fig8}  shows that the lateral pupil position measurement performance is in agreement with the end-to-end simulations and 10$\times$ better than the requirement.  The longitudinal pupil tracking performance could not be characterised at the MPE laboratory, because of the unavailability of longitudinal pupil actuators inside the GRAVITY instrument. It is tested at the VLTI laboratory using the variable curvature mirrors available in  the delay lines (cf. Section~\ref{sec:sky_char}).

\subsection{Field tracking  performance}

\begin{figure}
\parbox[t]{1pt}{\rotatebox{90}{\hspace{1cm}$|\alpha_{\rm in} - \alpha_{\rm out}|$ (mas)}} \hspace{5pt}
\includegraphics[width=0.2\textwidth]{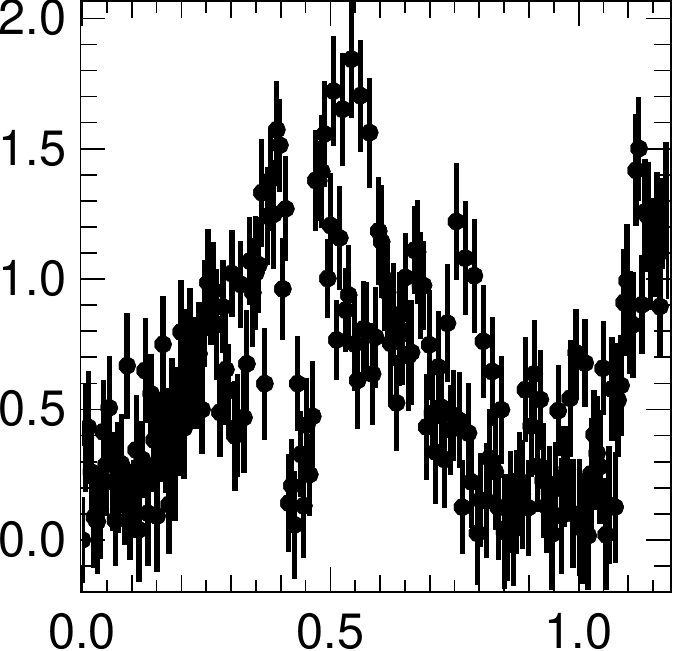} \hspace{1pt}\parbox[t]{1pt}{\rotatebox{90}{\hspace{1cm}$|\alpha_{\rm in} - \alpha_{\rm out}|$ (mas)}} \hspace{5pt}
\includegraphics[width=0.2\textwidth]{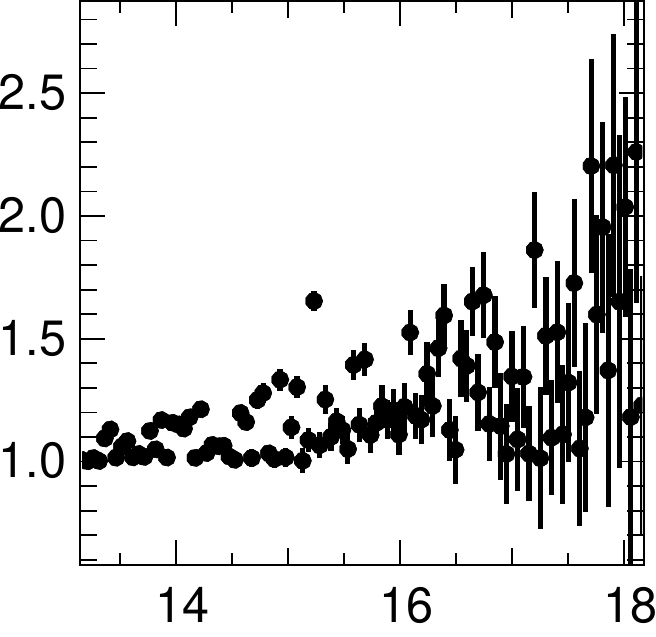} 

\hspace{1.6cm} $\alpha_{\rm in}$ (arcsec) \hspace{2cm} Magnitude, $m_\mathrm{H}$
\caption{Performance of the field tracker in the laboratory for UT setup. Left: The absolute error as a function of true object position. The observed shape is due to the mechanical instability of the tip-tilt correcting system. Right: The absolute error of object tracking as a function of star magnitude, $m_\mathrm{H}$. The error bars are RMS errors.} 
\label{fig:fig9}
\end{figure}

To characterize the performance of the field tracker, a known tip-tilts values $\alpha_\mathrm{in}$ are applied to the incoming beams and measured them back $\alpha_\mathrm{out}$ using  the field tracker as explained in Section~\ref{FImes}. The tip-tilts to the incoming beams are applied using the tip-tilt actuators. Figure~\ref{fig:fig9} left panel presents the absolute error of the object position as a function of a known input tilt. The error is less than 2\,mas. These measurements are done at high signal-to-noise conditions, i.e., negligible readout noise centroid error contribution. Figure~\ref{fig:fig9} right panel presents the absolute RMS error as a function of the magnitude of a star. An absolute RMS error of 1\,mas is observed for the equivalent of a $m_\mathrm{H}=13$\,mag star.

\begin{figure}
\parbox[t]{1pt}{\rotatebox{90}{\hspace{0.65cm} Normalized flux}}\hspace{5pt}
\centering
\hspace{3pt}
\includegraphics[width=0.45\columnwidth]{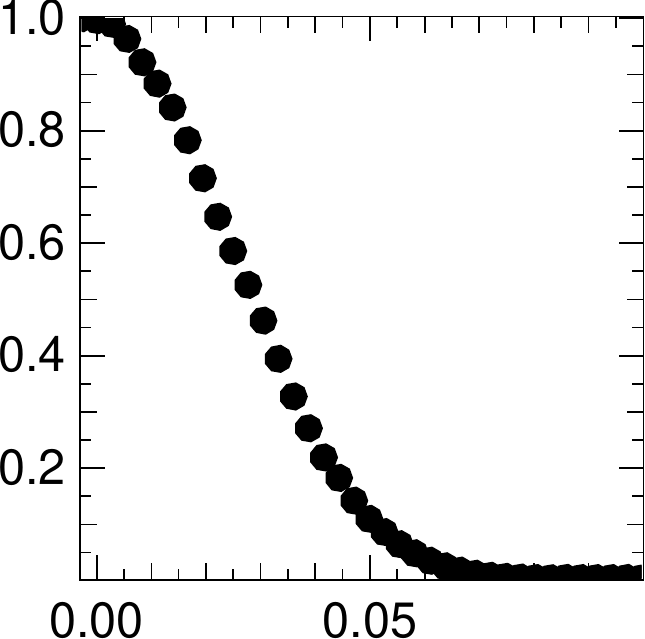} 

\hspace{1cm}Tilt ($^{\prime\prime}$)\caption{Fibre injection performance at the laboratory.} 
\label{fig:fig10}
\end{figure}

\paragraph*{Injection performance} Figure~\ref{fig:fig10} shows how the injection of star light into the fibre drops with unwanted tilts. From the figure, it can also be seen that the size of the fibre core  is around 60\,mas. Using the field tracker enabled field stabilization the coupling efficiency of the fibre is maintained around 75\%.

\paragraph*{Atmospheric differential refraction} The atmospheric differential refraction offset $\Delta R$ (Eq.~\ref{ADR_shift}) importance can be computed using the effective wavelengths and a zenith angle $\zeta=30\degree$. It is $\sim\,22$\,mas (which corresponds $\sim\,1.23$ pixel shift) and corresponds to a $\sim\,20$\% loss of injection of light into the fibre.

\subsection{Beam aberration sensing and  performance}

\begin{figure}
\parbox[t]{1pt}{\rotatebox{90}{\hspace{0.5cm} RMS wavefront error (nm)}}
\centering
\hspace{1pt}
\includegraphics[width=0.45\columnwidth]{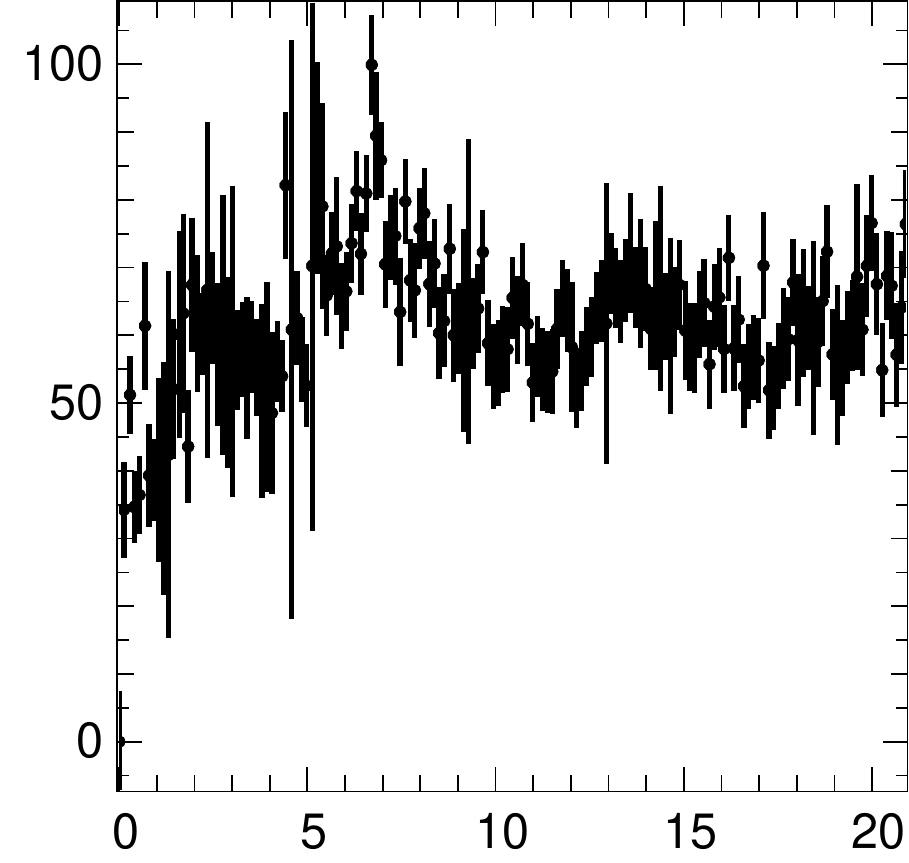} 

\hspace{1cm}Input wavefront error (at $\lambda_{\rm H}=1.65\micron$)\caption{The RMS wavefront error as a function of an input wavefront applied at the laboratory (UT beam). The error bars are RMS errors.} 
\label{fig:fig11}
\end{figure}

The aberration sensor wavefront estimation performance test is carried out in two steps. First, the focal lengths of the Shack-Hartmann lenslet are calibrated by applying series of tilts to the incoming plane wavefront~\citep{Kumar2013}. Second,  to measure the accuracy of the aberration sensor, known tip-tilt and defocus wavefront aberrations are applied to the incoming beams and measured them back using the aberration sensor function as explained in Section~\ref{ABmes}.  Figure~\ref{fig:fig11} presents the RMS wavefront error observed as a function of an input wavefront (tip, tilt and defocus RMS errors). It can be seen that the wavefront accuracy is $\approx \lambda/20$ at the UT scale. In this experiment a sub-aperture flux equivalent to a $m_\mathrm{H}=11$\,mag star ($\sim 2500\,\mathrm{ADU}\,\mathrm{s}^{-1}$) is used. 

\section{On-sky characterization}\label{sec:sky_char}

\subsection{Image quality characterisation}

\begin{figure}
\parbox[t]{1pt}{\rotatebox{90}{\hspace{-2.1cm}meter}} 
\parbox[t]{1pt}{\rotatebox{90}{\hspace{-6.5cm}arcsec}} 
 
\centering
\includegraphics[width=0.45\textwidth]{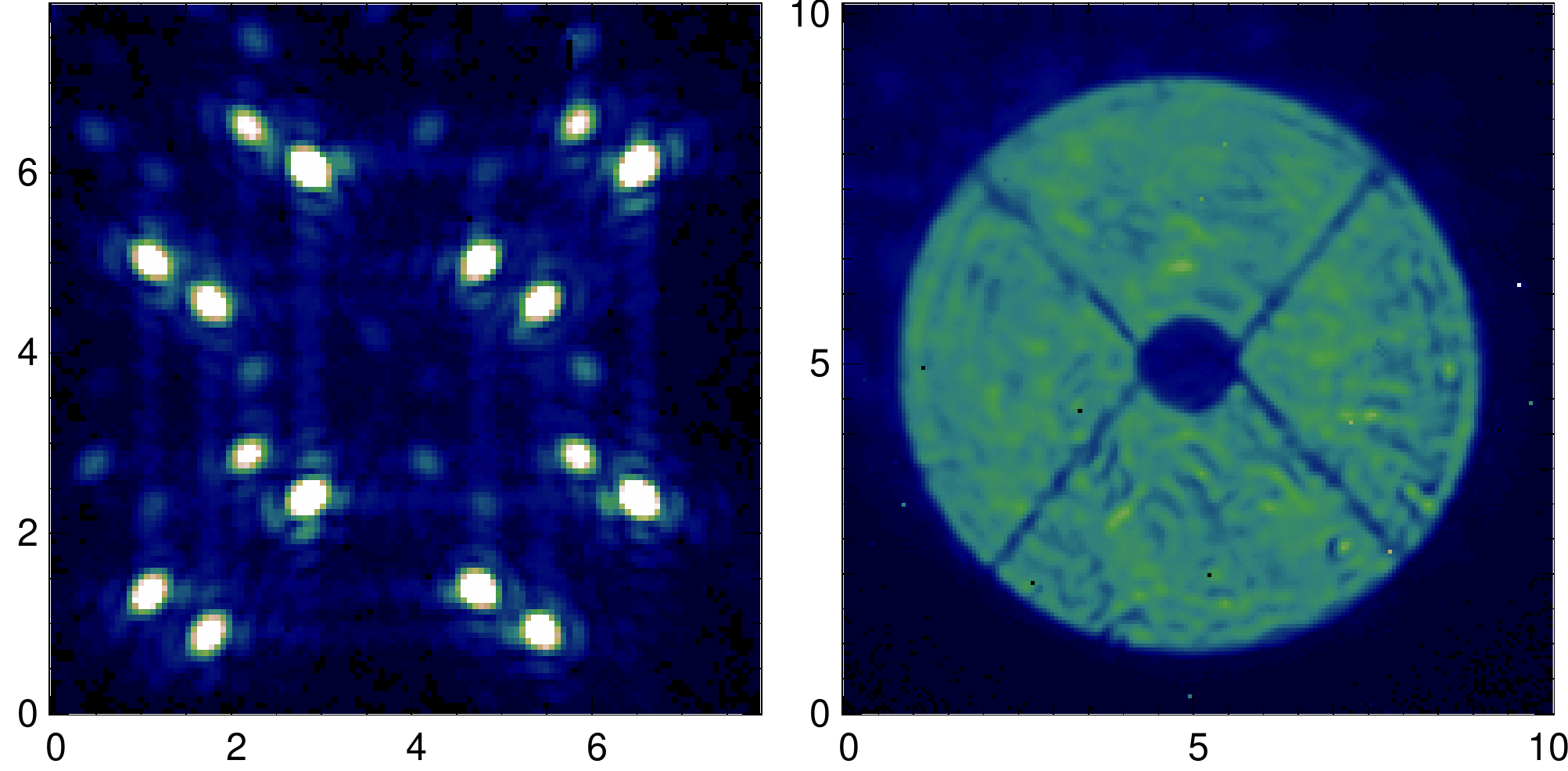} meter  \hspace{3cm} meter

\vspace{3pt}
\def\svgwidth{0.92\columnwidth}
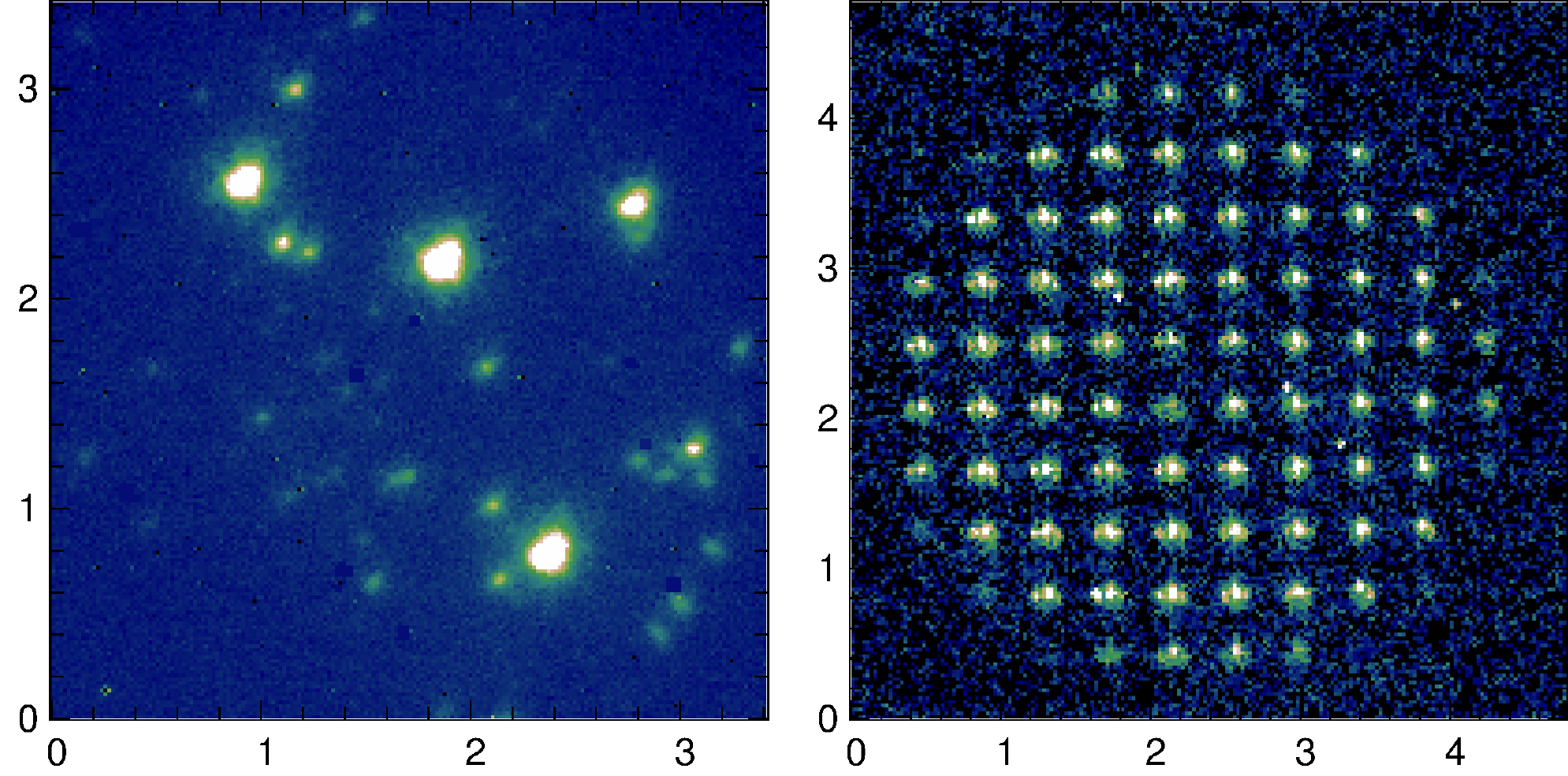 arcsec \hspace{3cm} arcsec

\caption{Imaging modes of camera observed on-sky with the UTs. Top (left to right): the  pupil tracker and the pupil imager. The low intensity cross stripes visible in the pupil imager are the secondary mirror holding spiders. Bottom (left to right):  the field tracker and the aberration sensor. The astrophysical target is the Galactic Centre.  The aberration sensor image is rotated counterclockwise to the field image due to a mirror reflection ($\pi$ phase). } 
\label{fig:fig12} 
\end{figure}

Based on verification tests at MPE, the GRAVITY instrument passed Preliminary Acceptance Europe (PAE) in May 2015. The instrument was then deployed at the Paranal Observatory, Chile and got its first light in November 2015. Figure~\ref{fig:fig12} presents the multiple beam acquisition and guiding camera detector image acquired on-sky with the UTs. Table~\ref{table:OnSkyCha} presents the Strehl ratio and the star FWHM at the camera when it was illuminated using the calibration unit. Also, the pixel scale which is measured, at the laboratory  and at the Paranal Observatory on-sky,  are presented.

\begin{table}
	\centering
	\caption{The field imager Strehl ratio, FWHM and pixel scales as built (at the UT scale).}
		\label{table:OnSkyCha}
	 \begin{tabular}{ | p{1cm} | p{1.2cm} | p{0.8cm} |  p{1.8cm} | p{1.8cm} |}
        \hline
        Telescope &Strehl&FWHM & Pixel scale &Pixel scale\\
        arm &ratio lab &at lab &at lab & on-sky\\
        &&(pixel)&(mas pixel$^{-1}$)&(mas pixel$^{-1}$)\\
        
    	\hline
		T1 & 0.90 & 2.65 &   18.23 &  17.77 \\
		T2 & 0.85 & 3.12 &   18.51 &  18.00 \\
		T3 & 0.70 & 3.18 &   18.20 &  17.64 \\
		T4 & 0.63 & 3.51 &  18.76 &  19.53\\
		\hline
	\end{tabular} 
\end{table}

\begin{figure}
\centering     
\includegraphics[width=0.45\columnwidth]{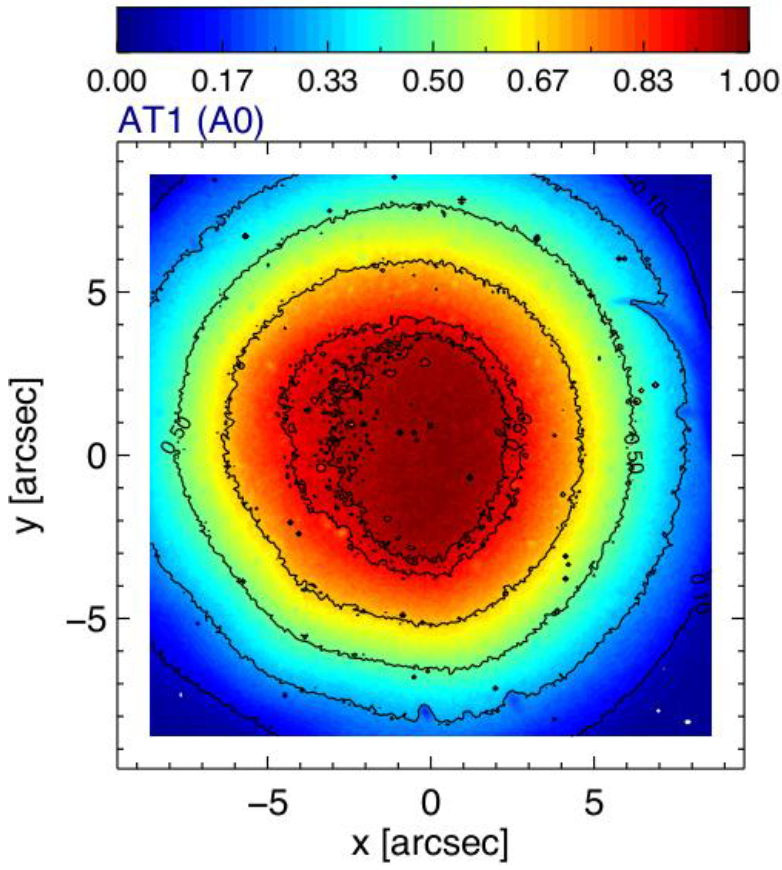}   \includegraphics[width=0.45\columnwidth]{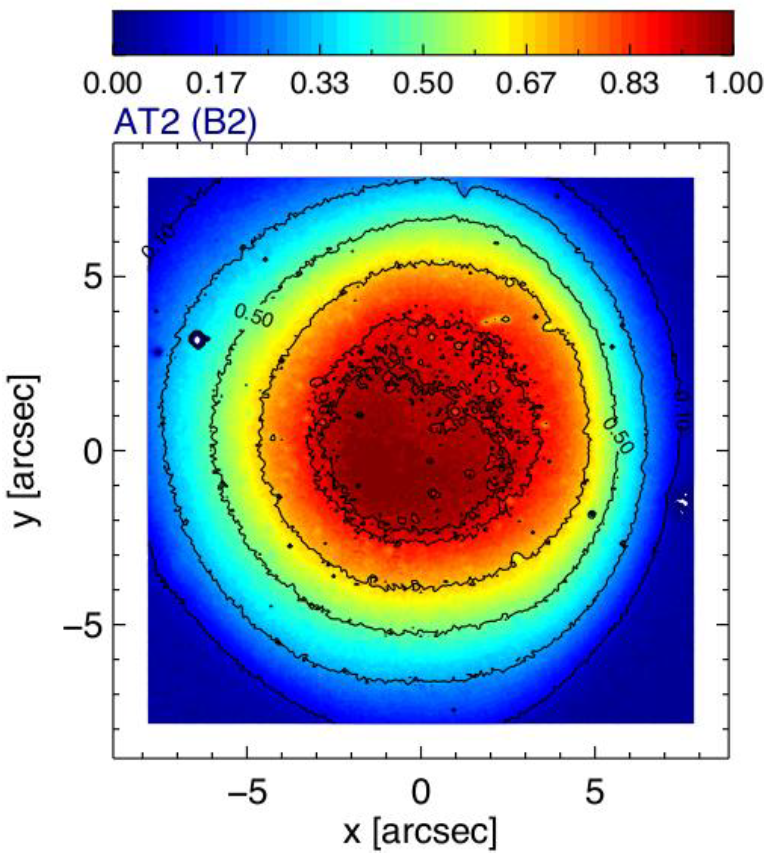}  
\caption{Representative fields of view of the field imager measured at two AT stations. The colour represents the normalised flux.}
\label{fig:FoVField}
\end{figure}

The sky field of view of the camera is larger than the nominal VLTI field of view ($2^{\prime\prime} \times 2^{\prime\prime}$). This is confirmed by  imaging (long exposures) the sky on the field imager and by moving the VLTI star separator tip-tilt mirrors~\citep[FSM,][]{Delplancke2004} in several steps.  Figure~\ref{fig:FoVField} presents the VLTI field of view and is around $3.6^{\prime\prime} \times 3.6^{\prime\prime}$ at the UTs. This value is less than the design value $4^{\prime\prime} \times 4^{\prime\prime}$ (cf. Table \ref{table:FISpecs}).

\subsection{The pupil tracker characterisation}
\begin{figure}
\parbox[t]{1pt}{\rotatebox{90}{\hspace{1cm}   $L_{\rm {z0}} - L_{\rm z}$ (mm)    }}\hspace{2pt}
\centering
\hspace{3pt}
\includegraphics[width=0.45\columnwidth]{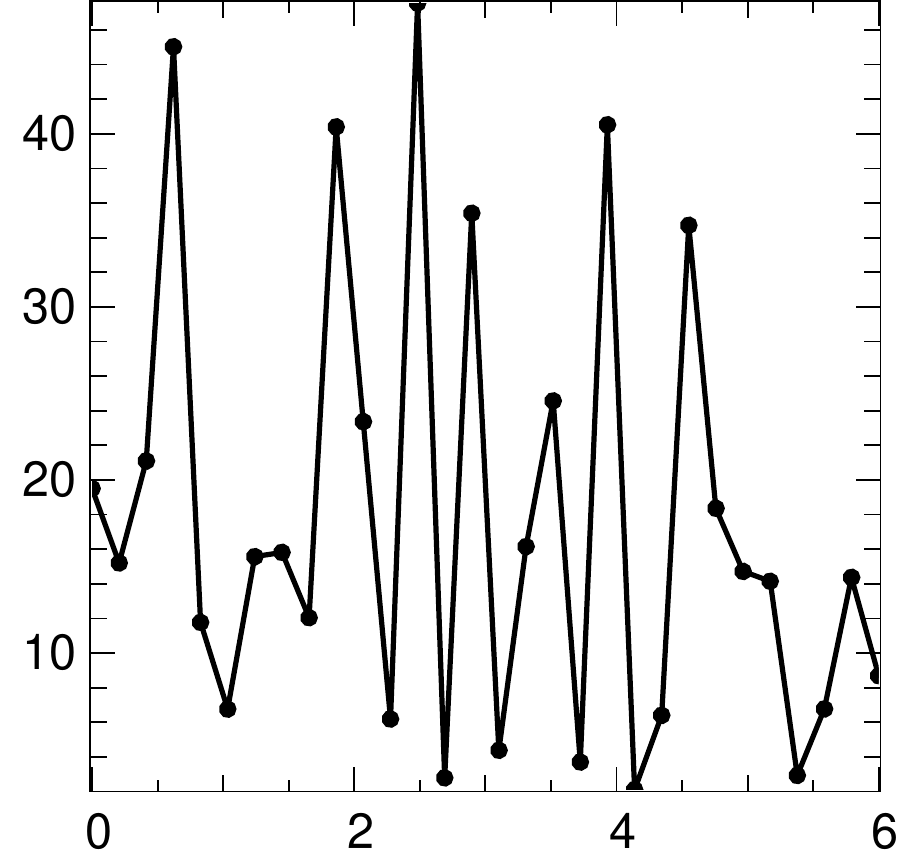} 

\hspace{1cm}$L_{\rm z0}$ (m) \caption{The longitudinal pupil shift residuals as a function of input longitudinal pupil shift ($L_{\rm z0}$) at the VLTI, for an UT beam.} 
\label{fig:fig13}
\end{figure}

\begin{figure}
\parbox[t]{11pt}{\rotatebox{90}{\hspace{-2.3cm}meter}} 
    
\includegraphics[width=0.45\columnwidth]{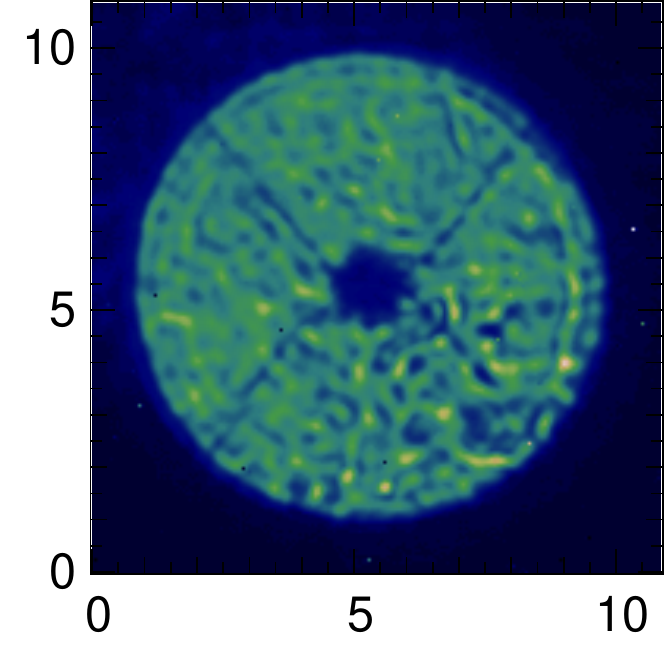}  \parbox[t]{1pt}{\rotatebox{90}{\hspace{1.5cm} meter}} 
 \includegraphics[width=0.45\columnwidth]{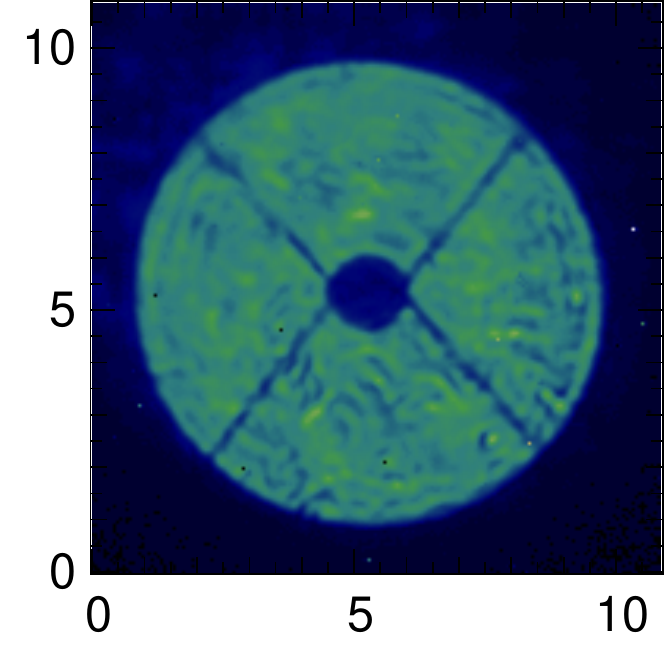}  
 
\hspace{2cm} meter \hspace{3cm} meter     \caption{UT pupil images  are obtained before (left) and after (right) closing the   pupil guiding loop.  }
\label{fig:fig14}
\end{figure}

Longitudinal pupil tracking performance is characterized at the VLTI laboratory by injecting artificial light source beams into the GRAVITY instrument via the delay lines, using the ARAL facility~\citep{Morel2004}. The longitudinal pupil shifts are simulated by manipulating the delay line variable curvature positions. Known longitudinal pupil shifts ($L_{\rm z0}$) are applied to the beam by moving the variable curvature mirror and the input shifts are measured back ($L_{\rm z}$) using the pupil tracker function as explained in Section~\ref{sec:PTmes}.  The longitudinal pupil accuracy (see Figure\,\ref{fig:fig13}) is better than 400\,m (i.e., $10^4$ magnification between the 8\,m telescope and 80\,mm beam inside the lab). Figure~\ref{fig:fig14} presents the UT pupil, before and after closing the pupil guiding loop. After closing the pupil guiding loop, an improvement of pupil focus can be seen. 

\begin{figure}
\parbox[t]{11pt}{\rotatebox{90}{\hspace{1cm}$L_{\rm x}$ or $L_{\rm y}$ (cm)}} 
\includegraphics[width=0.4\columnwidth]{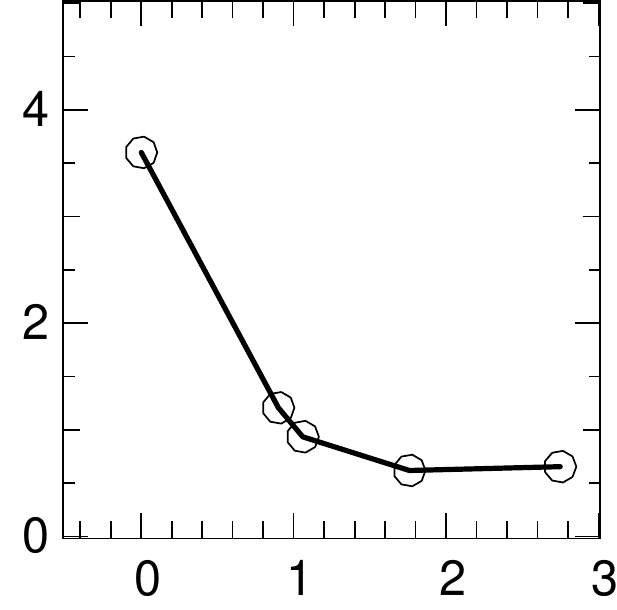}  \parbox[t]{1pt}{\rotatebox{90}{\hspace{1.5cm}$L_{\rm z}$ (m)}}  		
\includegraphics[width=0.45\columnwidth]{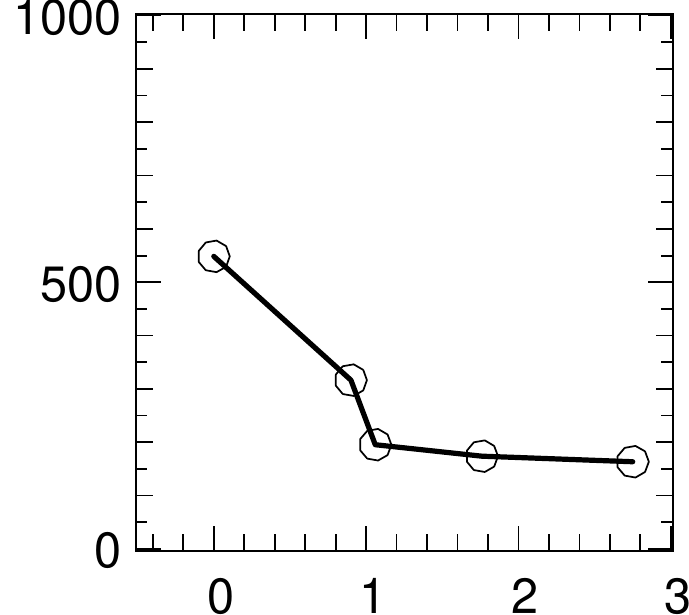} 

\hspace{1.4cm}Magnitude, $m_\mathrm{H}$ \hspace{1.8cm} Magnitude, $m_\mathrm{H}$

\caption{Pupil tracker performance (scaled to the UTs) as a function of the magnitude of several target stars. Left to right panels are the lateral pupil residuals along X or Y directions and the longitudinal pupil residuals.}
 \label{fig:fig15}
\end{figure}

\paragraph*{Bright target performance}  The pupil tracker experiences high backgrounds from bright  astrophysical targets observed in the J and H-bands due to: a) the closeness of the operating wavelengths of the pupil tracker (1.2~$\micron$) and the field imager; b) because of no adequate filter is available for the pupil tracker. The best operation of the pupil guiding is possible when: a) the flux of the telescope pupil reference laser beacons is high; b) imaging a lower magnitude or red coloured astrophysical target (with low flux at 1.2\,$\micron$). The background problem is solved to a certain extent on two fronts: a) using the neutral density attenuation filters\footnote{In order to not saturate the camera detector in bright astrophysical targets, GRAVITY is equipped with H-band neutral density attenuation filters (10 and 5\,magnitudes) in  front of the camera.}; b) using the BLINK mode.  This mode removes the background by taking two frames with the laser beacons on and off. The exposure with the beacons on is subtracted from an exposure with the beacons off. This effectively removes the background caused by the astrophysical target. With this mode, the performance of the pupil guiding is improved additionally by 2 magnitudes. The performance of the pupil guiding as a function of target magnitude is presented in Figure~\ref{fig:fig15}.  The pupil tracker performance degrades significantly for targets brighter than $m_\mathrm{H}=1$\,mag.

\subsection{The pupil imager}

The pupil imager is, currently, used to visually monitor the telescope pupils for their quality during the preparation for observations and also while observing.  However, it can also be used for the lateral pupil tracking in the presence of a bright astrophysical source  to complement the problem of the pupil tracker (see Section~\ref{PImes}). 

\subsection{The aberration sensor focus calibration}

\begin{figure}
\centering     
\parbox[t]{11pt}{\rotatebox{90}{\hspace{0.8cm} RMS defocus  (in $\lambda_{\rm H})$}} 
\includegraphics[width=0.4\columnwidth]{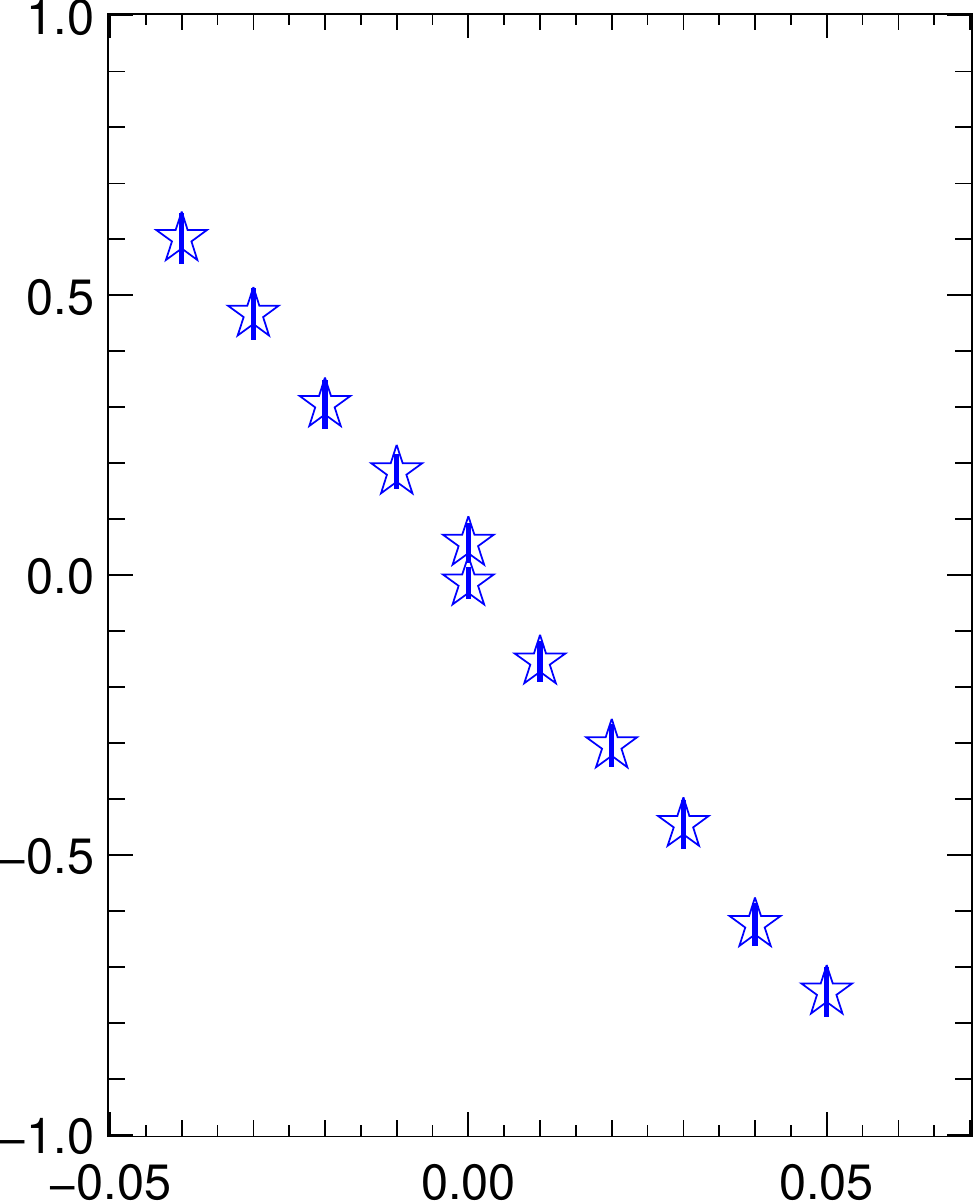}\hspace{11pt}
\parbox[t]{11pt}{\rotatebox{90}{\hspace{0.8cm}  FWHM of star (pixel) }} 
\includegraphics[width=0.4\columnwidth]{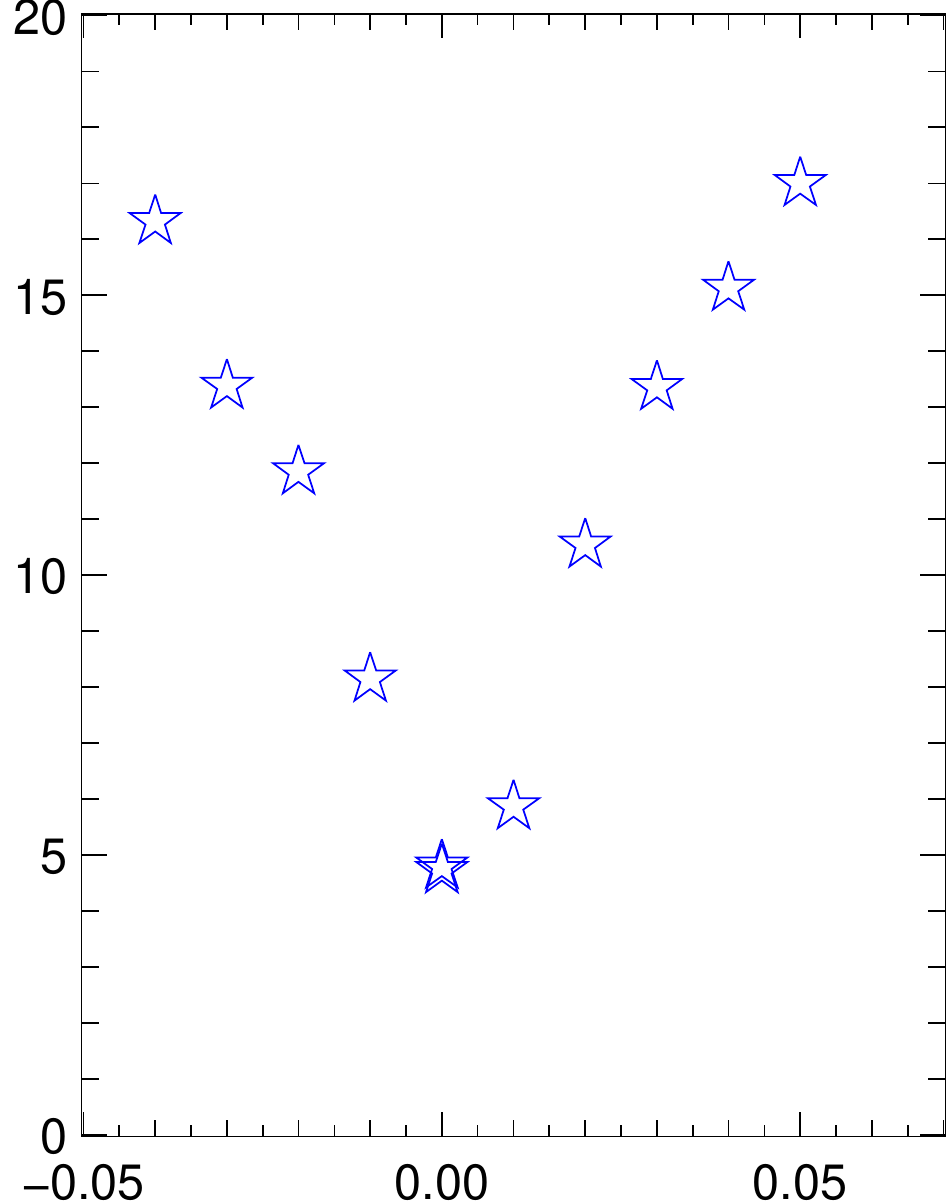} 

\hspace{1cm} M2 position (a.u.) \hspace{1cm}  M2 position (a.u.)
\caption{Focusing calibration of the ATs with the aberration sensor. Left: Zernike defocus coefficient RMS versus M2 focus position (in arbitrary units -- a.u.).  Right: The corresponding FWHM of a star observed in the field imager.}
\label{fig:ABSMes}
\end{figure}

The aberration sensor is used for the focus correction for ATs. Figure~\ref{fig:ABSMes} shows the aberration sensor focus measurement calibration as a function of  the secondary mirror (M2) position (in arbitrary units).  The right panel presents the corresponding star FWHM measured in the field imager.

\subsection{The field and pupil guiding residuals}

The field and pupil beam stabilizations are achieved with two types of actuators. The smaller field and pupil offsets occurring during interferometric observations are corrected in the closed loop using the GRAVITY internal actuators for speed and accuracy.  Larger field and pupil offsets, which usually occur during the initial alignment of GRAVITY, are corrected using the VLTI  based star separator actuators and delay line variable curvature mirror. If when operating in closed loop the offsets are larger than internal actuators range, they  are offloaded to the VLTI star separator actuators. 

\begin{figure}
\parbox[t]{1pt}{\rotatebox{90}{\hspace{1cm}  Field residuals (mas)   }}\hspace{3pt}
\includegraphics[width=0.95\columnwidth]{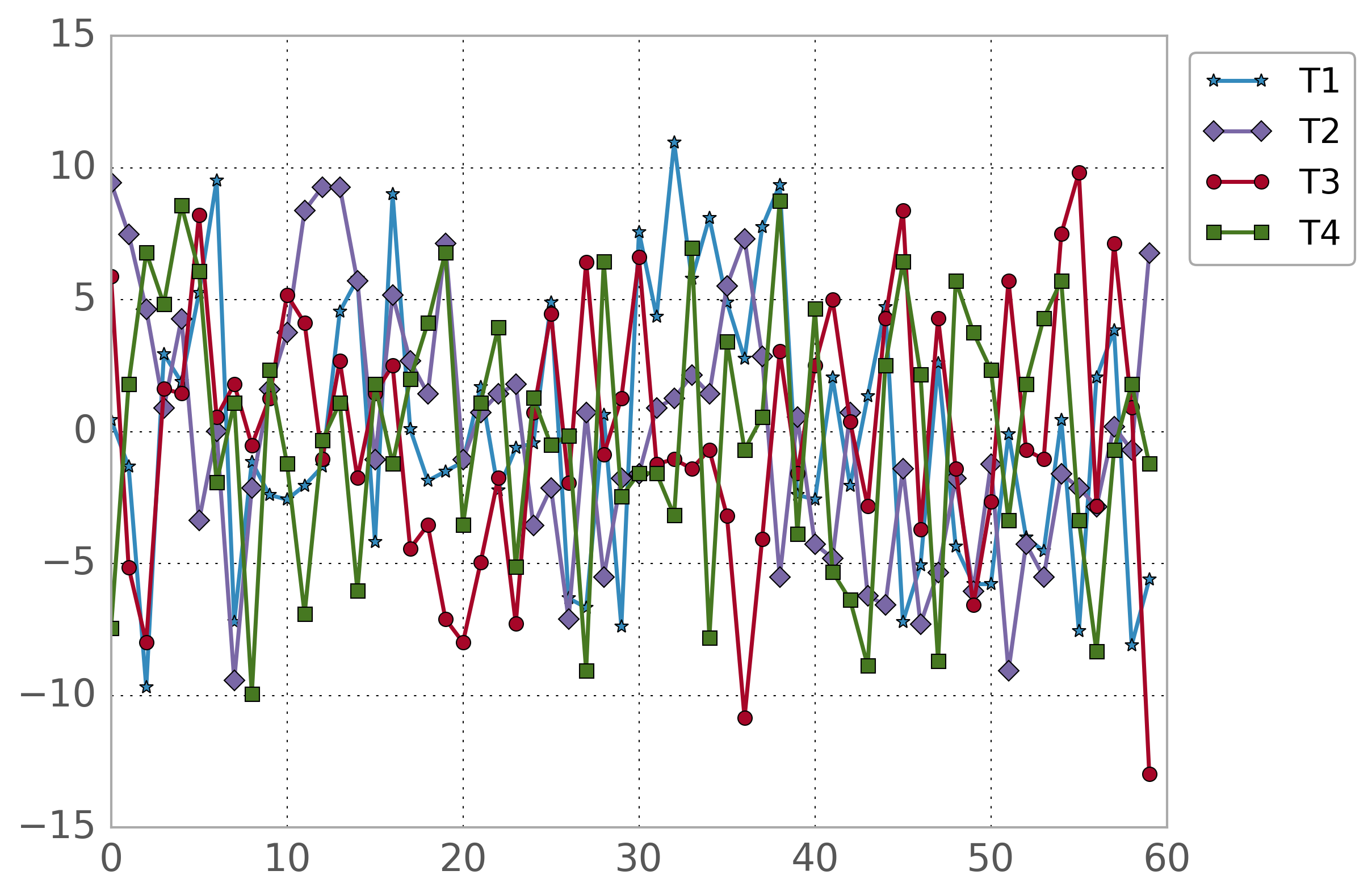}       

\hspace{2.5cm}  Continuous frames    \caption{The field guiding residuals for object HD 141742 ($m_\mathrm{H}=6.17$) when observed with the UTs on 15$^\mathrm{th}$ September 2016. A neutral density filter of 5 magnitudes was used.}
\label{fig:FIResidual}
\end{figure}

\begin{figure}
\parbox[t]{1pt}{\rotatebox{90}{\hspace{0.2cm}  Lat. residuals (mm)   }}\hspace{5pt}
\includegraphics[width=0.95\columnwidth]{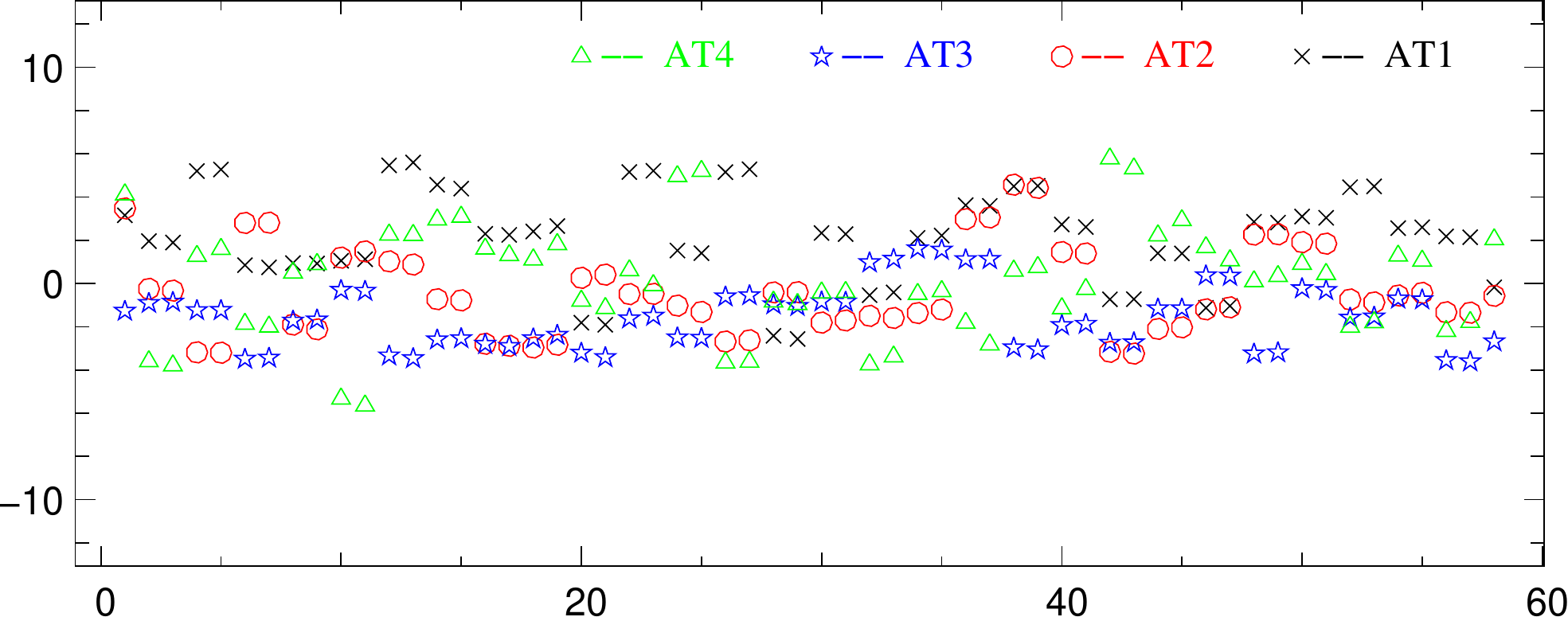} 

\hspace{1cm} Number of measurements (for each second frame) 

\vspace{11pt}
\parbox[t]{1pt}{\rotatebox{90}{\hspace{0.2cm}  Long. residuals (m)   }}\hspace{5pt} 
\includegraphics[width=0.95\columnwidth]{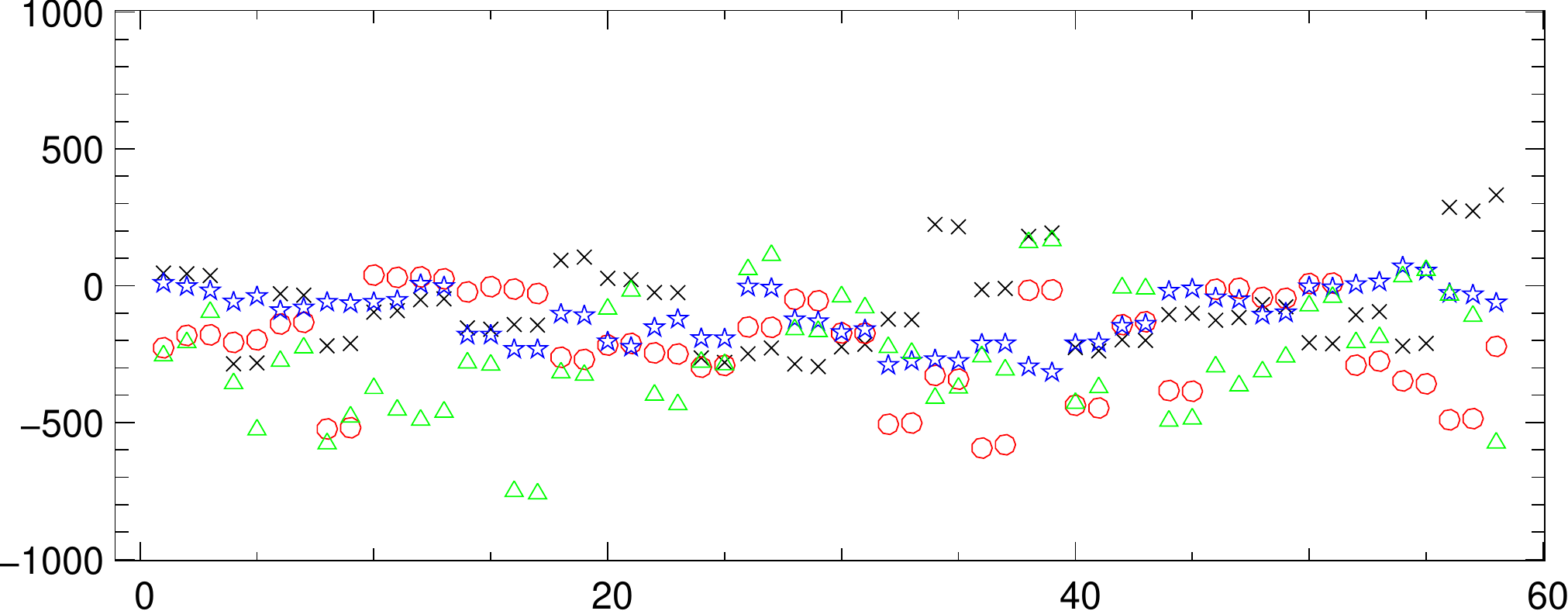}       

\hspace{1cm} Number of measurements (for each second frame) 
\caption{Pupil guiding residuals observed for the ATs, for 60 frames on  19$^\mathrm{th}$ March 2016. The measured pupil shifts are scaled to the UT scale. The astrophysical target has $m_\mathrm{H}=1.74$. The pupil tracker is operated in the BLINK mode during this experiment.}
\label{fig:PTResidual}
\end{figure}

Often, GRAVITY  experiences optical misalignments with the VLTI, in the worst case,  the field offsets of  $\sim 60$\,mas RMS  and lateral pupil offsets of  $\sim$5\% (0.4\,m RMS) in the UT pupil. With the multiple beam acquisition and guiding camera enabled beam guiding, the field, lateral pupil and longitudinal pupil guiding are implemented with standard deviation residuals smaller than $\pm 0.5$  pixels RMS (10\,mas RMS),  $\pm 0.2$ pixels RMS (12\,mm RMS) and $\pm 400$\,m RMS at the UT respectively. Figures~\ref{fig:FIResidual} and \ref{fig:PTResidual} present the field and pupil guiding residuals. 

\subsection{Astrometric residuals study}

\begin{table}
	\centering
	\caption{Astrometric error at the UT due to field and pupil residuals.}
	\label{table:AstroCalc}
    \begin{tabular}{ | p{1.5cm} | p{1.8cm} | p{2.2cm} |  p{1.5cm} |}
    \hline
    & Residual & One & Four \\ 
    &errors &telescope &telescopes\\
    \hline
    Without beam guiding residuals & $\Delta \alpha~\leq~60\,\mathrm{mas}$ $\Delta L_{\rm x}~\leq~0.4\,\mathrm{m}$  $\Delta L_{\rm z}~\leq~30$\,km  & $316.6\,\mu$as   $\delta OPD~\leq~153.5\,\mathrm{nm}$  & $\leq 633.3\,\mu$as  \\ 
    \\
    Beam guiding requirements  &  $\Delta \alpha~\leq~10\,\mathrm{mas}$ $\Delta L_{\rm x}~\leq~40\,\mathrm{mm}$ $\Delta L_{\rm z}~\leq~10\,\mathrm{km}$ &   4\,$\mu$as    $\delta OPD~\leq~1.95\,\mathrm{nm}$ & $\leq 8\,\mu$as\\
    \\
    Laboratory beam guiding residuals  &  $\Delta \alpha~\leq~2\,\mathrm{mas}$ $\Delta L_{\rm x}~\leq~4\,\mathrm{mm}$ $\Delta L_{\rm z}~\leq~400\,\mathrm{m}$ & ~~~~-- & ~~~~~-- \\ 
    \\
    On-sky beam guiding residuals  &  $\Delta \alpha~\leq~10\,\mathrm{mas}$ $\Delta L_{\rm x}~\leq~12\,\mathrm{mm}$ $\Delta L_{\rm z}~\leq~400\,\mathrm{m}$ &   1.2\,$\mu$as    $\delta OPD~\leq~0.58\,\mathrm{nm}$ & $\leq 2.4\,\mu$as\\ 
    \hline
    \end{tabular}
\end{table}

The  total astrometric error produced by the four telescopes (ref Eq.~\ref{sigma_T}) adds to

\begin{equation}\label{sigma_4T}
\sigma_\mathrm{4T} = \sqrt{ \sum_{i=1}^4\sigma_{\mathrm{T},i}^2}~.
\end{equation}

The astrometric error caused by the field and the pupil motions is around 633.3\,$\mu$as and 2.4\,$\mu$as before and after closing the beam guiding loop, respectively, at the UT and for the baseline of 100\,m. Table~\ref{table:AstroCalc} gives the summary of the guiding requirements and the achieved beam guiding is several conditions. The on-sky residuals are larger than the laboratory characterization (using the calibration unit generated beams) results because the multiple beam acquisition and guiding  camera frame rate (0.7\,s) is not quick enough to correct them.

A GRAVITY observation is described in detail in~\citet{GRAVITYCollaboration2017}. An animation of starlight tracing through the GRAVITY instrument is available elsewhere\footnote{\href{http://www.eso.org/public/videos/eso1622b/}{http://www.eso.org/public/videos/eso1622b/}}. 

Operation of GRAVITY with the multiple beam acquisition and guiding camera started with the ATs in October 2016 and with the UTs in April 2017. GRAVITY  first light and science verification results are published in, e.g., \citet{GRAVITYCollaboration2017, GRAVITYCollaboration2017A, GRAVITYCollaboration2017B, GRAVITYCollaboration2017C, LeBouquin2017, Kraus2017}. The first verification of the narrow-angle astrometry of GRAVITY is done with an M-dwarf binary, GJ 65 and estimated the binary separation with residuals of 50\,$\mu$as~\citep{GRAVITYCollaboration2017}. However, the astrometric residual errors can be reduced further by computing the guiding residual errors carefully from the multiple beam acquisition and guiding camera images in off-line in the data reduction pipeline and inputting them into the astrometric calculations. There are also other possible error contributors in the GRAVITY measurements but they are out of the scope of this paper to describe in detail:  a) uncertainty in the calibration of the narrow-angle astrometric baseline and wavelength; b) dispersion in the single-mode fibres and integrated optics.

\section{Summary and conclusions}

Accurate beam acquisition and guiding camera methods for  phase referencing optical/infrared interferometry are presented.  These methods offer advances in  near-infrared imaging and optical/infrared long-baseline interferometry: a) accurate and active pupil and field guiding required for sensitivity and astrometry; b) the characterization of the input telescope beams quality, the focus correction for the ATs  and the possibility of correcting the non-common path aberration errors between the adaptive optics system and GRAVITY. 

The beam acquisition and guiding camera performance is verified using laboratory-generated telescope beams and on-sky at the VLTI.  The characterization results show that it is able to analyse the telescope beams: a)  field tracking in a crowded field with $\lesssim 10$\,mas RMS residuals translates in a 75\% coupling efficiency of near-infrared starlight into single-mode fibres b) lateral pupil tracking with residuals less than 12\,mm RMS; c) longitudinal pupil tracking with residuals less than 400\,m RMS; d) quasi-static higher order wavefront aberration measurements with 80\,nm RMS. The multiple beam acquisition and guiding camera measured beam parameters are used in the stabilization of the field, pupil and the ATs focus correction.  The pupil imager is used to visually monitor the telescope pupils for their quality during the preparation for observations and also while observing.  In overall, the performance of the camera is in agreement with the requirements of the GRAVITY science programs, although some work is still needed to push the residual errors further down by implementing the post-processing of the multiple beam acquisition and guiding camera images in the data reduction pipeline and also by correcting the non-common path aberration errors between adaptive optics and GRAVITY. 

The adaptive optics assisted  camera imaging is  an asset for the GRAVITY science observations as it provides  H-band images of the target using four different telescopes, simultaneously, in parallel to the interferometric observations in  the K-band. This imaging can be useful in three ways: a) to extend wavelength coverage in the H-band (1.45-1.85\,$\micron$); b) to extend spatial resolution coverage (low resolution images from the field imager and the high resolution images from the interferometric large baselines); c) to solve the $2\pi$ interferometric phase wrapping ambiguity\footnote{If the two stars separated more than $\lambda/B$, they suffer phase $2\pi$ wrapping ambiguity.} using the multiple beam acquisition and guiding camera field images, provided that  the  camera is astrometrically calibrated \citep[e.g.][]{Yelda2010}.

Although the multiple beam acquisition and guiding camera is a subsystem of GRAVITY, it can be used for other interferometric beam combiner instruments available at the VLTI ~\citep[e.g. MATISSE,][]{Lopez2014} for the purpose of an automated alignment with the VLTI and for closed loop tip-tilt guiding. Furthermore, the accurate pupil guiding developed here is of relevance: a) to high-contrast imaging as it requires pupil stabilization enabled extreme adaptive optics ~\citep[e.g. SPHERE,][]{Montagnier2007}; b) to correct the misregistration of pupils in between the wavefront sensor and the deformable mirror, where they have moving optics in between\footnote{e.g., the case where the deformable mirror placed at the telescope or the Coud\'{e} focus and the wavefront sensor placed inside the beam combiner laboratory.}; c) to the accurate alignment of pupil mask to match the telescope pupil in order to limit the background radiation coming into the spectrographs \citep[e.g. HARMONI,][]{Hernandez2014}. The pupil tracker allows both the telescope pupil rotation and the lateral shifts estimation accurately. 



\section*{Acknowledgements}
We thank the technical, administrative and scientific staff of the participating institutes and the observatory for their extraordinary support during the development, installation and commissioning of GRAVITY.  We thank the referee for her/his constructive and insightful report.  Anugu acknowledges FCT and IDPASC-Portugal for his PhD grant SFRH/BD/52066/2012.   The research leading to these results received funding from PTDC/CTE-AST/116561/2010, COMPETE FCOMP-01-0124-FEDER-019965, UID/FIS/00099/2013. This research was partially funded by European Commission Seventh Framework Programme, under Grant Agreements 226604 and 312430 (OPTICON Fizeau Programme).






\bsp	
\label{lastpage}
\end{document}